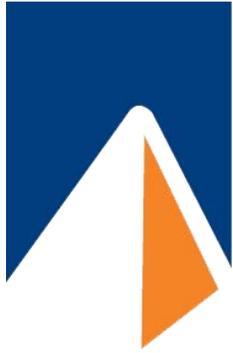

# Analysis of Ground Plane Size, Topography and Location on a Monopole Antenna's Performance Utilizing 3-D Printing

**Division of Engineering Programs**

**Master of Science Thesis**

**Student: Vito Ciraco**

**Supervisor: Dr. Reena Dahle**

Signature Page

Analysis of Ground Plane Size, Topography and Location on a Monopole Antenna's Performance Utilizing 3-D Printing

Vito Ciraco

State University of New York at New Paltz

We, the thesis committee for the above candidate for the Master of Science degree,

Hereby recommend acceptance of this thesis.

---

Professor Reena Dahle, Thesis Advisor

Division of Engineering Programs, School of Science and Engineering,

SUNY New Paltz, New York

---

Professor Ghader Eftekhari, Thesis Committee Member

Division of Engineering Programs, School of Science and Engineering,

SUNY New Paltz, New York

---

Professor Sima Noghanian, Thesis Committee Member

Principal Antenna Design Engineer at Commscope, San Jose, California

---

# Dedication

This thesis is dedicated to my beloved parents, Vito and Anna for their undying love and support to further my education. All thanks be to God for guidance, strength and health to complete this thesis.

# Acknowledgements

It is my genuine pleasure to express my deep sense of thanks and gratitude to my advisor and mentor, Dr. Dahle for her continuous support, patience, motivation, enthusiasm, and immense knowledge. It has been my honor to be advised and mentored since the beginning of my college career to the dissertation of my master thesis by such a wonderful Professor. I want to thank the entire the faculty at the Division of Engineering Programs at SUNY New Paltz for making my education possible. I offer my sincere appreciation for the learning opportunities provided by the committee members as well as their patience and guidance. Special thanks should be given to Spiro Argyros, Katherine Wilson and Matthew Pilek as well as my colleague Issa Nesheiwat.

# Abstract

The monopole antenna is widely used in communication applications and is typically mounted on various surfaces that act as ground planes; a prime example being the roof of a car. The shape of the ground plane can drastically change the patterns of the electromagnetic radiation of a monopole antenna as well as its RF performance. Extensive work [1,12-13] has been done on the numerical modeling of arbitrarily shaped ground planes. However, due to their geometric complexity, there is very little work reported on the practical testing component of physical antennas with these obscure ground plane structures. This thesis illustrates how the additive manufacturing process presented can be used to physically realize arbitrarily shaped ground planes and provides a low-cost process to verify the numerical model. Ground Planes were modified while maintaining the same antenna length to evaluate the impact on antenna performance. The antenna was not optimized or changed to a standard antenna design. Varying radius spherical ground planes are modelled, as well as modified ground plane structures to evaluate the impact of the ground plane on a 1.3GHz monopole antenna's performance and in some cases to modify the antenna's performance in terms of gain, bandwidth, and radiation pattern. Designs such as the planar ground with horn was found to enhance monopole bandwidth by more than 5 times that of a standard planar ground but significantly deteriorate the antenna's radiation pattern. Moreover, complex geometry such as the fin sphere ground plane offered a 25% increase in gain relative to the standard sphere ground. Designs like the edge-mounted sphere can offer directive gain and radiation characteristics simply by altering the antennas' location mount location with respect to its ground plane. The techniques presented in this thesis offer new ways of producing 3-D printed ground planes for RF applications that are easier to manufacture, lighter in weight, and can enhance antenna performance over their conventional counterparts.

# Content

# Lists of Tables

# List of Figures

# Chapter 1

# Introduction

## 1.1 Motivation

Wireless communication has been incorporated in everyday devices as technology and humanity are constantly progressing. Most of these devices today utilize some type of antenna to transmit information wirelessly. Common examples of wireless vehicles today are cars, planes, or aerospace probe applications such as satellites and spacecrafts. When an antenna is mounted on a vehicle of limited physical size, the ground plane reference for the antenna will also be limited. Ideally, an antenna would have an infinitely large, flat, perfectly conducting ground plane. There has been work reported on how ground plane size can affect antenna characteristics and performance [1-3], however, not much research has been done on how to create complex 3-D structure ground planes and alter them to enhance antenna performance. Additive manufacturing has enabled the creation of antennas and other RF components rapidly and at a low cost [4-5] compared to conventional manufacturing methods. 3-D printing allows for production of creative and complex shapes that would otherwise be difficult to, or impossible to manufacture by conventional means.

## 1.2 Scope

In this thesis, a monopole antenna with an operating frequency of 1.3GHz is mounted to various simple and altered ground planes. The ground planes are manufactured using a 3-D printer with Acrylonitrile Butadiene Styrene (ABS) as the structural material, which then undergoes multiple post-processing metallization techniques.

Two basic geometries were chosen as the shapes of the ground planes under test: a sphere and a flat plane. Multiple sizes of radius for the ground planes are examined as well as alterations to the topography of the ground planes. Extensive electromagnetic (EM) simulations are conducted to gauge the performance of antennas theoretically prior to manufacturing. Once modelled, the antennas and ground planes were fabricated, assembled, and tested to compare to the simulation results. A standard monopole antenna with an ideal, flat ground plane was also modelled and tested as a reference to verify the simulated model is accurate to theorical results.

## 1.3 Overview

Chapter 2 discusses the background of how antennas operate and the different types of antennas. The fundamentals of 3-D printing and how it has been used for RF applications is also presented in Chapter 2. Chapter 3 discusses the fabrication and assembly process of the antenna and ground planes. Next, the different designs of the altered planar and spherical ground planes are presented in Chapter 4. This chapter also details how these different designs were simulated and optimized. Chapter 5 presents the measured data obtained from the manufactured designs in comparison to the simulations. Finally, Chapter 6 presents the concluding remarks and discusses future proposed work.

# Chapter 2

## Background Theory

### 2.1 Introduction

The Institute of Electrical and Electronics Engineers (IEEE) Standard Definitions of Terms [7] defines the antenna as "a means for radiation or receiving radio raves". An antenna is a transitional structure between free space and a guiding device [6]. More simply, an antenna is the interface between electric currents moving in a conductor and electromagnetic waves propagating through space. This electrical current is either fed from a transmitter to the antenna to send radio waves out into space or from the antenna taking in radio waves (inducing an electric current) into a receiver. Figure 2.1 depicts an antenna as a transitional device between free space and a guiding device (a transmission line).

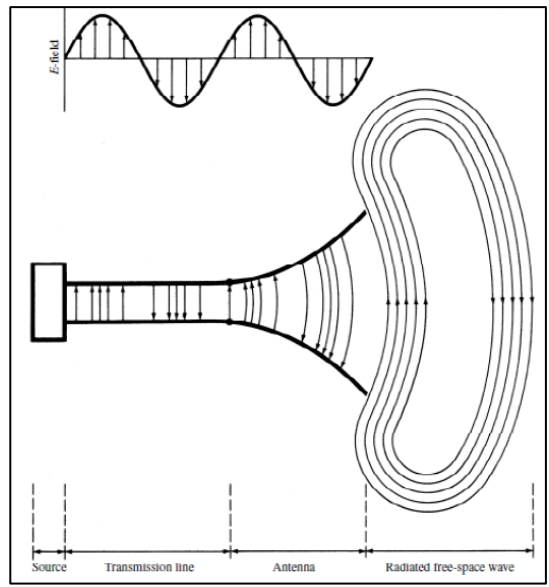

**Figure 2.1:** Antenna shown as a transitional device [6].

Antennas produce electromagnetic waves whenever a varying current is applied. When current is passing through a conductor, electric and magnetic fields are produced around the wire. If the current is varying within the wire, these electric and magnetic fields also vary while moving out from their source, projecting electromagnetic waves out into space.

## 2.2 Types of Antennas

There are many kinds of antennas that take diverse form factors to accomplish different performance characteristics for the application it is designed for.

### 2.2.1 Wire Antenna

The most prevalent antenna is the wire antenna and can be seen frequently in daily life because of their simplicity and low cost. They are commonly found on cars, buildings, ships, handheld radios and much more. Wire antennas come in many different forms, most commonly as a monopole, dipole, circular loop, or helix, as shown in Figure 2.2 [6]**.**

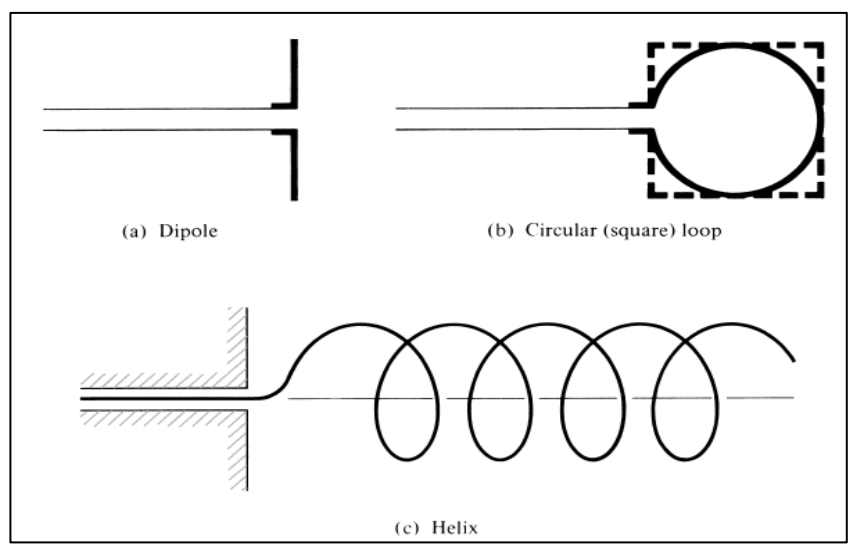

**Figure 2.2:** Common wire antenna configurations [6].

The antenna being studied in this thesis is a monopole wire antenna. It is very similar to the dipole antenna shown above in Figure 2.2, however, it consists of only one wire instead of two. A ground plane is implemented in replacement of the second wire to make a functional monopole antenna. In its simplest form, a monopole antenna above an infinite ground plane can be considered half of a corresponding double-length, center fed linear dipole antenna [14]. The current distribution along the wire is a standing wave when resonating. This occurs because the ground plane creates an image of the monopole with a current distribution identical to that for the lower arm of the dipole. Figure 2.3 shows a vertical monopole above a ground along with the corresponding center-fed dipole [14]. Effectively, the ground plane acts as the other half of the dipole but does not radiate into space. All the radiation is directing into the upper half of space, creating a power density that is twice as high as that for a dipole radiating the same power [14].

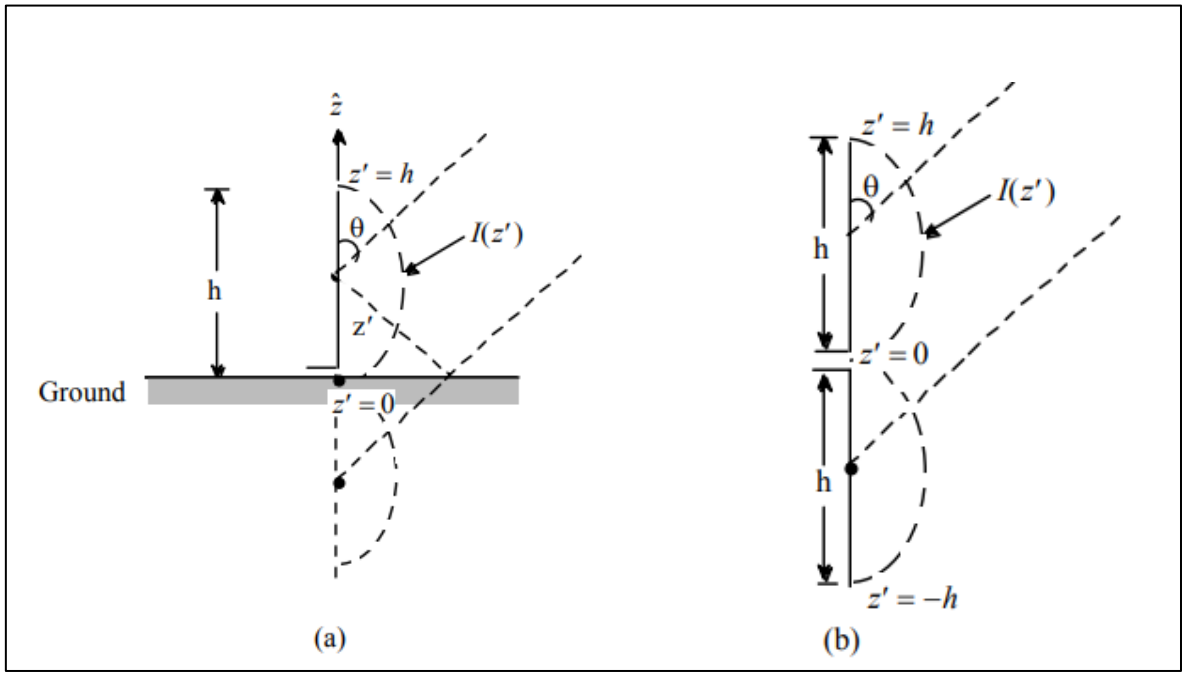

**Figure 2.3**: (a) Vertical monopole antenna above ground, and (b) corresponding center-fed dipole [14].

## 2.3 Ground Planes

A ground in electrical terms is the reference point in a circuit; the ground carries a voltage of zero. It is a large conducting surface in comparison to the wavelength and serves as a reflecting surface for waves. A large, infinite ground plane is required and ideal to have a strong, uneasily disturbed point of reference for an antenna. However, this is not always realizable, especially on surfaces where the footprint is being minimized. The ground plane shape and size play major roles in determining the antenna's characteristics including impedance, radiation pattern and gain.

### 2.3.1 Infinite Ideal Ground Plane

Ideally, a ground for a linear wire antenna is an infinitely large, perfectly conducting plane with the wire antenna perpendicular to the ground plane. The antenna could be arranged in a multitude of ways, including mounting horizontally or at angle. More complex monopole designs include a folded monopole or a parasitic monopole [14]. These more compact geometries are lower profile but are restricted by the presence of the conducting ground. The radiating elements require a quarter wavelength distance between the antenna element and its ground plane to avoid unwanted image currents. If too close to the ground plane, radiation from the image currents will destructively interfere with the direct radiation from the antenna [15]. The wire antenna is typically mounted vertically, perpendicular to the ground plane as can be seen in Figure 2.4, depicting a vertical dipole with generated field lines.

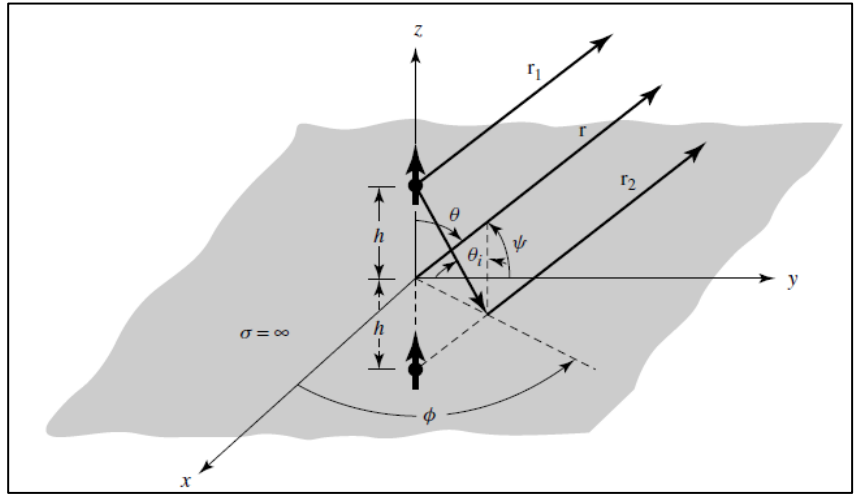

**Figure 2.4:** Vertical dipole perpendicular to ground with radiating fields [6].

Time varying current on the ground plane is the primary source of radiation that determines both the antenna's impedance and radiation pattern [2]. A ground plane with a radius larger than $10\lambda$ will have a small effect on impedance; finite size ground planes have significant effect on both radiation and impedance [15]. For example, a quarter wavelength monopole mounted on a cell phone has a measured gain of 1dBi, while the same antenna mounted on an infinite ground would have a gain of 5dBi. This large drop in gain is attributed to some energy being radiated in the lower half of space due to the small size of the ground plane [15]. A good conducting ground extends to a radius of at least equal length of the antenna element, preferably larger by 20-50% [15]. The ground plane for a given antenna has a direct impact on the fields generated. The total field above the ground plane is equal to the sum of the direct and reflected components of the electric field [6]. The equation to find the far electric field direct component of the monopole of length $L$ and constant current $I_0$ is shown in equation (2-1), while the equation to find the reflected component is shown in (2-2).

$$E_\theta^d = j\eta \frac{kI_o h e^{-jkr_1}}{4\pi r_1} \sin(\theta_1) \qquad (2\text{-}1)$$

$$E_\theta^r = jR_v\eta \frac{kI_o h e^{-jkr_2}}{4\pi r_2} \sin(\theta_2) \qquad (2\text{-}2)$$

Since the field cannot exist within the perfectly conducting ground plane, the electric field is equal to 0 below the surface. The origin of the coordinate system can be moved to the surface of the ground plane (z=0) to allow simplification of the expression for the total electric field. Vectors $r_1$ and $r_2$ can be found by equations (2-3) and (2-4) below [6],

$$r_1 = [r^2 + h^2 - 2rh \cos\theta]^{1/2} \qquad (2\text{-}3)$$

$$r_2 = [r^2 + h^2 - 2rh \cos(\pi - \theta)]^{1/2} \qquad (2\text{-}4)$$

For far field observation, where vector r is much larger than the length of the antenna h, equations (2-3) and (2-4) can be reduced using binomial expansion to the equations shown below in (2-5) and (2-6).

$$r_1 \approx r - h \cos\theta \qquad (2\text{-}5)$$

$$r_2 \approx r + h\cos\theta \qquad (2\text{-}6)$$

Finally, utilizing Equations (2-5) and the sum of Equations (2-1) and (2-2), a general equation for the total electric field for a vertical monopole on a flat infinite ground plane is shown in (2-7).

$$E_\theta \approx j\eta \, \frac{kI_o h e^{-jkr}}{4\pi r} \sin\theta [2\cos(kh\cos\theta)] \quad z \geq 0$$

$$E_\theta = 0 \quad\quad\quad z < 0$$

(2-7)

Theoretically, a monopole mounted vertically above a flat perfect conductor produces an E-Plane radiation pattern like the one presented in Figure 2.5 [6]. The radiation pattern from the antenna mounted on a flat infinite ground creates 2 symmetric half-circular lobes. However, practical, finite ground planes can introduce unwanted back radiation lobes from the small ground plane radiating some energy from the induced currents [16].

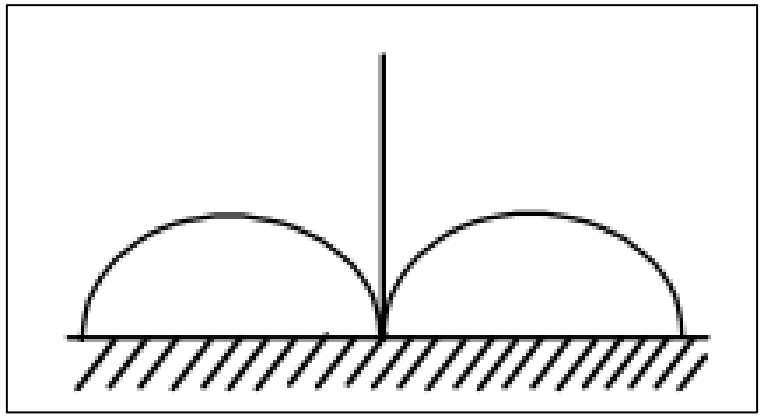

**Figure 2.5:** Theoretical E-field radiation pattern for a monopole mounted vertically on a flat perfect conductor [14].

### 2.3.2 Spherical Ground Planes

A time varying current on the ground plane is the primary source of radiation that determines both the antenna's impedance and radiation pattern [2]. If the ground plane itself is curved, there is a precipitous drop in radiated power density at the curved surface [14]. For example, the curvature of the Earth must be considered even when utilizing planet Earth as the ground when radiating at distances more than 1000 miles, because the radiated power drops considerably on the surface as it curves [14]. There has been work reported on how curved ground planes affect antenna performance [8-9,12]. Extensive research has been conducted on dielectric resonator antennas (DRA), due to their characteristically high radiation efficiency and shape flexibility. A spherical ground plane structure with a hemispherical DRA mounted to it has been reported in [8] and is depicted in Figure 2.6. The metallic ground sphere is assumed to be a perfect conductor and the results of this report were of numerical modelling only. The DRA has an operating frequency of 1.7GHz. The result of utilizing the spherical ground over a planar ground resulted in a decrease in the input resistance and reactance. Furthermore, the resonant frequency slightly increased with the increasing radius of the ground plane sphere [8]. Figure 2.7 displays graphical results of the input resistance, input reactance and reflection coefficient from this paper.

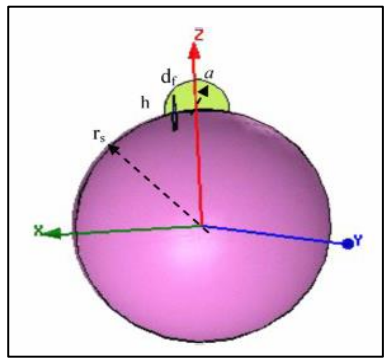

**Figure 2.6:** Hemisphere DRA mounted on a spherical ground plane [8].

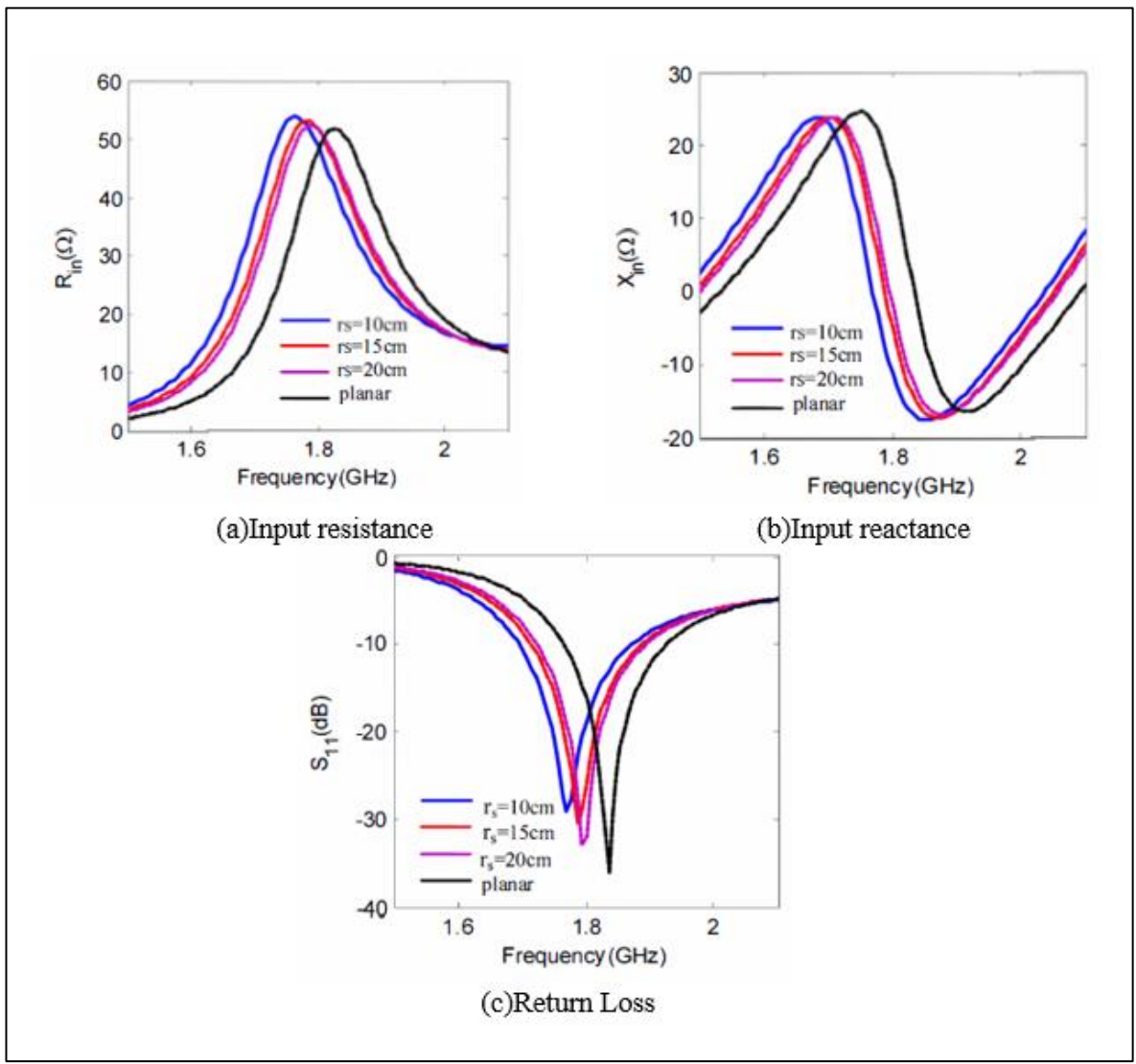

**Figure 2.7:** (a) Input resistance, (b) input reactance and (c) reflection coefficient in terms of sphere radius [8].

2.3.2 Curved Ground Planes

High impedance surfaces (HIS) have been utilized for gain and bandwidth enhancement, surface wave suppression and especially useful for low profile antenna designs [9]. The ground plane reported in [9] utilizes a series of rings separating copper patches on a curved disk substrate to create the HIS as depicted in Figure 2.8 with a blue highlighted loop antenna. The spherical disk ground is made up of ABS plastic that is then painted with conductive silver paint. This ground plane has a curvilinear antenna mounted to it that is resonant at 3GHz. The results of this report concluded that the curved surface caused a 10% decrease in bandwidth with a loop antenna and 16% decrease in bandwidth with a spiral antenna. The gain on the non-broadside of the ground plane suffered a decrease in bandwidth, while the broadside gain was 3dB higher than the flat ground HIS [9]. The radiation pattern found for the curved HIS ground with a loop antenna can be seen in figure 2.9 in comparison to a flat ground plane where the curved HIS has considerable larger radiation intensity.

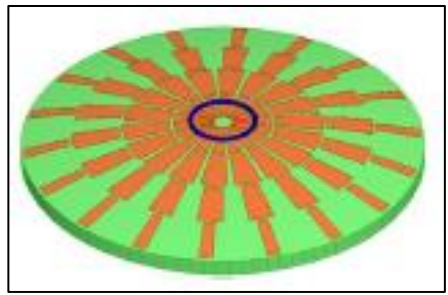

**Figure 2.8:** Geometry of circular HIS with spherical curvature [9].

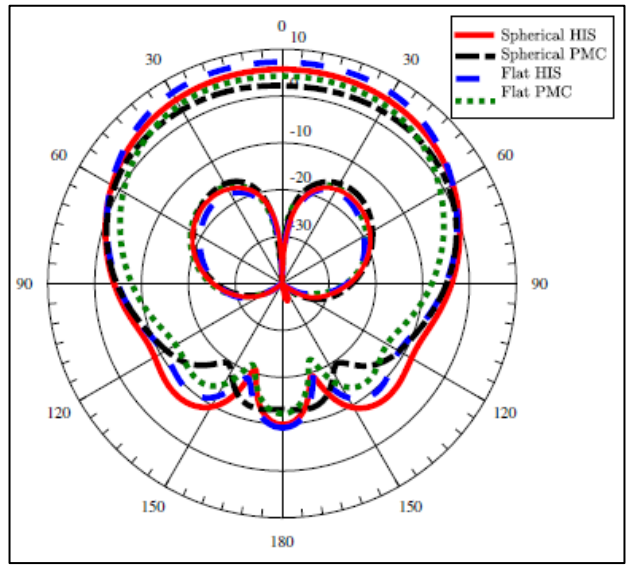

**Figure 2.9**: Radiation pattern of loop antenna on HIS curved ground compared to a planar ground [9].

In [17], the effects on performance of bending the ground plane on a mobile terminal antenna is presented. The mobile antenna is a non-resonant type, capacitive coupling element (CCE) which can be seen in Figure 2.10 mounted to a planar ground plane. The ground plane is 50mm in width, and either 105mm or 175mm in length. The CCE is used to couple efficiently to the wave modes on the ground plane since the resonance at the desired frequency is obtained using an external matching circuit [17]. This enables more flexible frequency performance since the antenna geometry is not used to obtain resonance, compared to a typical self-resonant type of antenna. The ground plane undergoes various radii of bending that ranges from being totally planar to 20mm in radius. Figure 2.11 depicts a profile view all the bending cases tested.

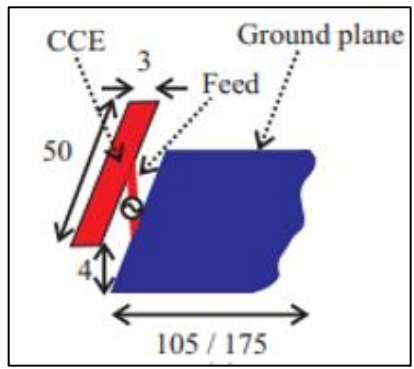

**Figure 2.10:** CCE antenna mounted to planar ground to be bent [17].

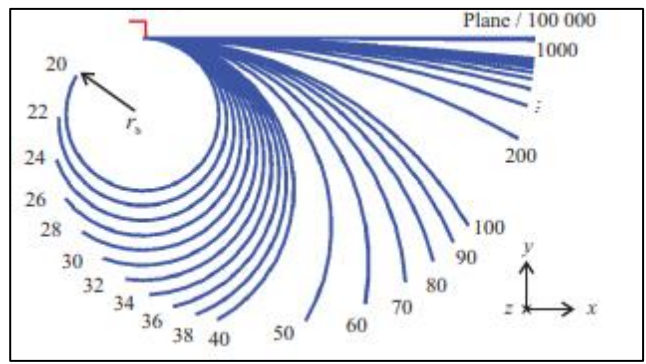

**Figure 2.11:** Profile view of bending cases for ground plane [17].

This paper found that the bending of the ground had a considerable effect on the impedance and bandwidth depending on the frequency and ground plane size. Due to the CCE characteristics, this antenna has a wide frequency of operation and different amounts of bending aided performance at some frequencies and hindered performance at others. The return loss along with the Smith chart for notable bending radii can be seen in Figure 2.12. It is evident that the 20mm bending radius enhanced the return loss at 1.2GHz by three times compared to no bend (planar) and 50mm radius bend.

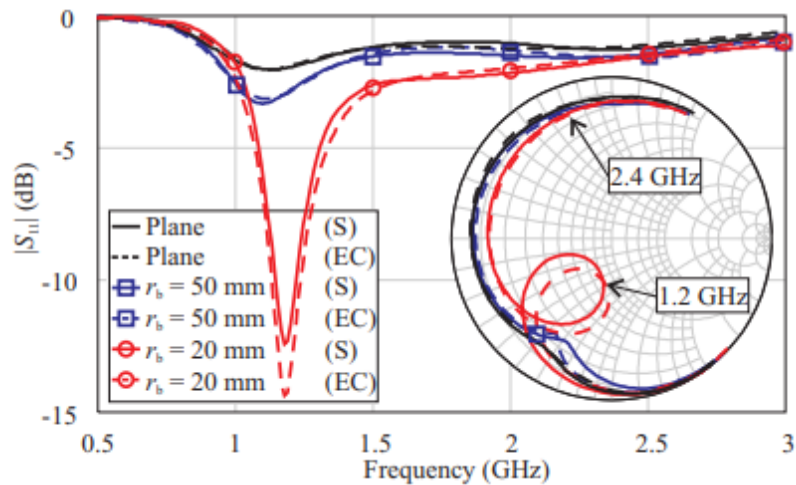

**Figure 2.12:** Return loss and Smith chart of CCE antenna with different radii of bending of the ground plane. [17].

## 2.4 3-D Printing

3-D printing technology has been growing in popularity due to its versatility and capabilities. 3-D printing is an additive manufacturing process in which a 3-dimensional object can be created in a matter of hours from a single digital file. There are a variety of different styles of 3-D printers that have emerged as the technology has progressed. The applications for 3-D printing are constantly growing as technology becomes more accessible.

## 2.4.1 Fused Deposition Modeling (FDM)

The most common style of 3-D printer today uses a method of additive manufacturing known as fused deposition modeling (FDM) due to its low cost. Material that can be used in FDM can range from simple strong plastics like Acrylonitrile Butadiene Styrene (ABS), to conductive filament for prototyping circuits. The 3-D printer heats an extruder to a precise temperature at which the material or filament becomes a liquid, that is then pushed through a nozzle to be deposited on a build plate where it cools and solidifies. This extruder can move precisely in 3-D space within the build area of the printer to deposit the melted material. This is done repeatedly, layer by layer, until the 3D object is formed. A diagram depicting how an FDM printer deposits material can be seen below in Figure 2.13. This type of 3-D printers has a standard layer resolution of 0.2mm per layer, and this can be modified by changing the nozzle size.

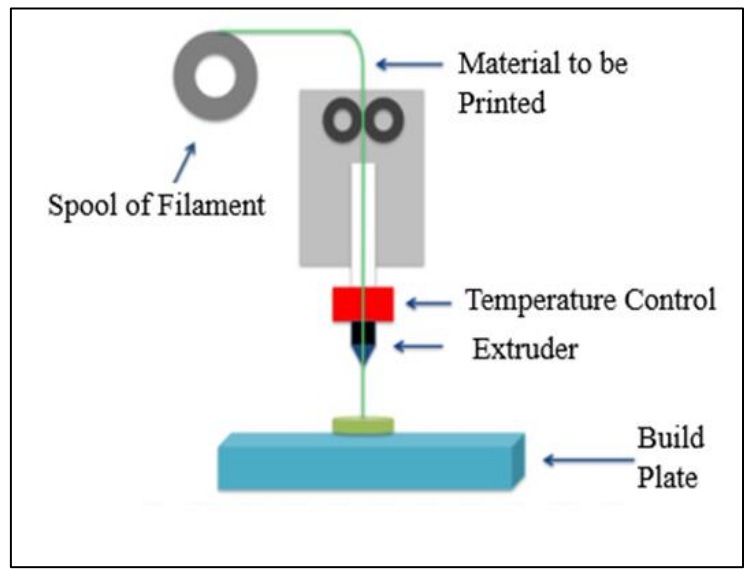

**Figure 2.13:** FDM style 3-D printing [10].

2.4.2 RF Applications of 3-D Printing

There are a multitude of applications for 3-D printing within the field of radio frequency engineering. Characterization of the dielectric properties of common plastics have been reported from different style printers [11]. Table 2-1 displays the dielectric constants and loss tangents found at 7GHz by a process of cavity perturbation measurement [11]. Upon evaluating the dielectric properties of multiple materials being printed from an FDM and stereolithography (SLA) style printer, it was found that the FDM style presented the lowest dielectric losses compared to SLA style with the same material. Therefore, FDM 3-D printing of materials has adequate dielectric properties to properly design microwave band devices such as antennas [11].

**Table 2-1:** Dielectric constants and loss tangent for different 3-D printed materials [11].

| Material reference | @ 7 GHz | |
| --- | --- | --- |
| | $\varepsilon'$ | $\tan\delta$ |
| TW-CON175BK | - | - |
| Ultem 9085 | 2.71 | $3.4 \cdot 10^{-3}$ |
| P430 | 2.39 | $3.6 \cdot 10^{-3}$ |
| M30 | 2.46 | $1.0 \cdot 10^{-2}$ |
| Vero Blue | 2.95 | $1.9 \cdot 10^{-2}$ |
| PPSF | 2.94 | $6.3 \cdot 10^{-3}$ |
| PC/ABS | 2.49 | $4.0 \cdot 10^{-3}$ |
| Polycarbonate | 2.57 | $3.8 \cdot 10^{-3}$ |
| Accura Xtreme | 3.00 | $2.9 \cdot 10^{-2}$ |

Applications of 3-D printed RF devices include electromagnetic absorbers, such as a fully 3-D printed termination. In [11], a fully 3-D printed termination for a WR-90 waveguide antenna is manufactured. This 3-D printed termination provided a VSWR lower than 1.025 over the entirety of the X-band [11]. Important advantages of 3-D printed RF components include cost and weight reduction; plastic is usually lighter in weight compared to metal. This can greatly reduce the weight of components, which can be especially useful for space probes or satellites.

3-D printing is also commonly used to fabricate dielectric antennas such as DRAs and lens antennas [18]. Lens antennas are shaped in such a manner to bend and refract radio waves traveling through it, in the same way an optical lens refracts light. 3-D printed lens antennas have been successfully demonstrated along a wide range of frequencies from the microwave band up to THz [18]. 3-D printing also allows for customized designs that can be easily and quickly manufactured. For example, seen below in Figure 2.14 is a 3-D printed 8-beam dielectric lens antenna that was designed for military mobile networks. The lens antenna shown can select between 8 different directional radiation beams or use all the beams at once for 360° coverage [18]. This lens was printed using multiple materials to take advantage of different dielectric properties. Attempting to manufacture an octagonal shaped, multi-material object by conventional techniques would be costly and difficult.

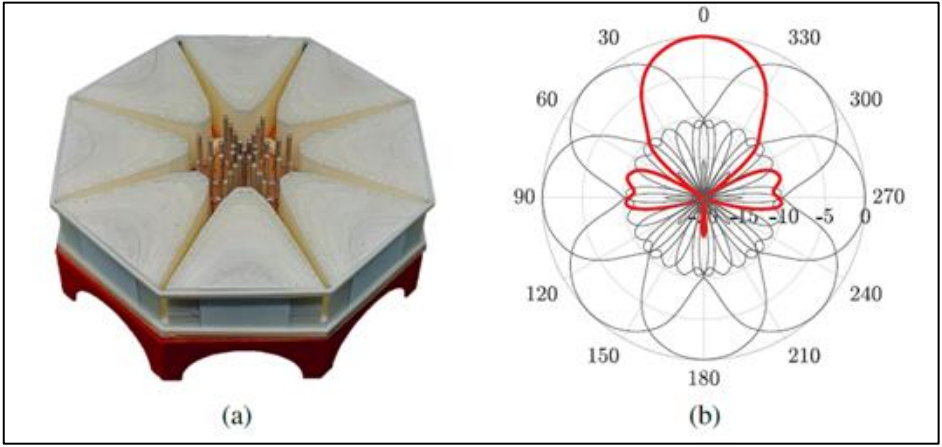

**Figure 2.14:** (a) 3-D Printed 8 beam dielectric lens antenna, (b) radiation pattern of 8 beam antenna [18].

Furthermore, the surface of a 3-D printed surface can be metalized to provide even better RF characteristics than just the 3-D printed material alone [10]. Metallization can be performed in a multitude of ways. For instance, spraying particles of metal within a special conductive paint is a simple, easy method. However, the metal layer added needs to be a minimum thickness of the skin depth for the signal. High frequency signals undergo the skin effect where the current travels only on the surface or skin of the conductor. The skin depth is dependent on the frequency of operation. It may be hard to achieve this thickness of metal with paint alone. More effective methods of metallization can be more complex, such as electroplating, which can allow for more metal to be deposited on the surface of the part. Electroplating is the process used in this thesis.

When using the 3-D printing process a challenge that rises is that the object's surface may not be perfectly smooth, there may be small ridges due to the multiple layers deposited. Smooth, clean surfaces produce the best RF characteristics, as well as aid in even distribution of metallization (if applicable). There are techniques to smooth the surface such as sanding or buffing, as well as chemical techniques which are discussed in greater detail along with electroplating in the next section

# Chapter 3

## 3-D Printing and Electroplating Process

### 3.1 3-D Printer

The 3-D printer utilized in this thesis is an upgraded version of the successful Ultimaker 2 platform [19], shown in Figure 3.1. This robust FDM style 3-D printer has been upgraded beyond its factory limits. Instead of the stock motherboard from Ultimaker, an open source 120MHz, 32 Bit board has been installed. The board is the SKR 1.4 Turbo from BIGTREETECH. Mounted onto the motherboard are 5 TMC2209 stepper motor drivers for accurate driving of the stepper motors that carry the extruder carriage and build plate assembly. Furthermore, the factory extruder has been removed and replaced with the top quality E3D v6 full metal hot end. Due to all these upgrades to the vital components and careful calibration, this 3-D printer can print complex shapes very accurately and uniformly. This printer has a build volume of 192mm x 205mm x 200mm.

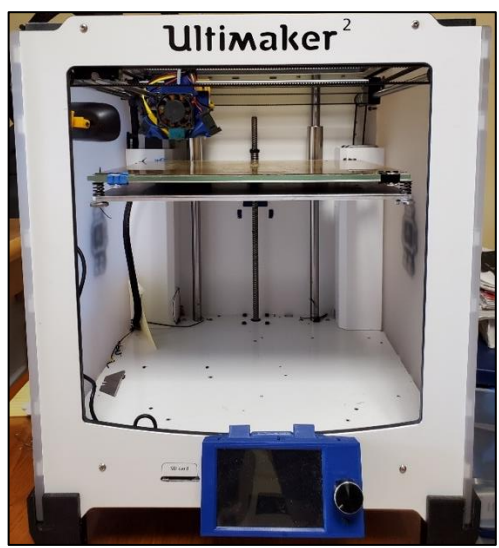

**Figure 3.1:** Ultimaker 3-D printer.

### 3.1.1 Printer Settings

For this thesis, the 3-D printer was fed a 1.75mm WyzWorks ABS filament. The dielectric constant and loss tangent found for this filament were 1.7 and 1.8E-2 respectively at 30% infill using a ring resonator setup. The extruder was heated to 250°C while the heated bed was heated to 90°C to allow better adhesion between the object and the build plate. ABS filament is known to warp during the printing process as the part cools overtime. This can result in a failed or misshapen print. To combat warping, glue stick was applied to the print bed prior to printing where the object will encounter the bed.

The nozzle diameter of which the plastic is extruded has a diameter of 0.4mm. The layer height was set to 0.2mm, which is a multiple of the diameter of the nozzle. This is known to produce the most uniform layers of material deposited. So, the 3-D printer will build each object layer by layer, of which each layer is 0.2mm in size along the z-axis.

The printer also generates support material when an object has steep overhangs or unsupported areas. For example, a support is needed if the printer is printing a simple archway, when the printer reaches the top of the arch, the center may require support material because there would be nothing but air beneath it to deposit filament onto. So, the printer generates a thin (material saving) scaffolding that will enable the print to deposit material successfully and the support material can be easily removed post printing. However, sometimes the object will have a rough surface where the support material contacts the object. This is worth noting as this thesis uses models that required support material, and a smoothing process needs to be applied where the support material was located.

### 3.1.2 ABS Smoothing

A smooth finish to the 3-D printed part is critical for RF applications and to assist in the successful metallization of the surface and reduce conductive losses. Due to the shapes of the models tested in this thesis, support material is also required which further degrades the surface of the 3-D printed part. There are post print processing techniques that can help smooth the surface of prints. The most common and simplest technique of 3-D print smoothing is sanding of the surface with sandpaper. Depending on the roughness of the surface, a certain level of grit sandpaper is chosen and polished across the surface of the part repeatedly. Lower numerical grit numbers have a rough surface while the higher the number, it is smoother. The objective of this is to use a small number grit sandpaper to smooth down especially rough surfaces, and then work up to a higher grit number to provide a smoother finish. This works for 3-D prints but tends to permanently damage the surface if performed repeatedly.

There is an alternative chemical technique of smoothing prints known as Acetone vapor smoothing. Acetone is a cheap and safe household chemical that is an effective removal solvent for a variety of polymers, including ABS. The 3-D printed part to be smoothed is suspended in a closed container with acetone placed in a basin on the bottom of the container. The time it takes for the acetone to smooth the surface varies and is dependent on the surface area of the print, and the humidity of the workspace. Figure 3.2 depicts a sphere directly after 3-D printing with support scarring visible and is compared to after ABS acetone smoothing the same 3-D printed sphere.

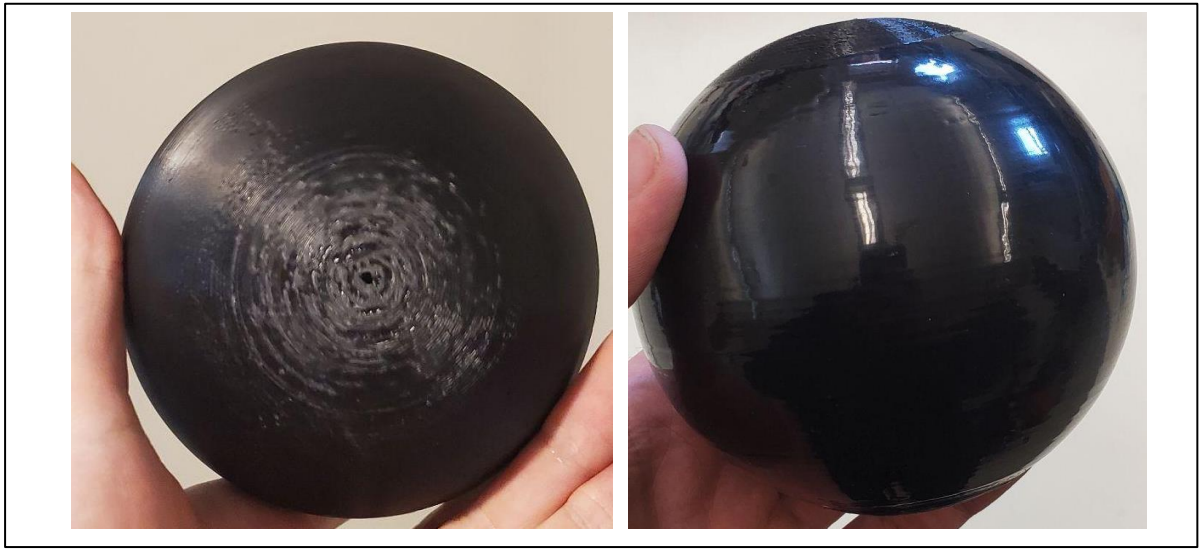

(a) Before acetone  (b) After acetone

**Figure 3.2:** 3-D Printed ABS sphere (a) before Acetone smoothing, and (b) ABS sphere after Acetone smoothing.

## 3.2 Electroplating

The process of electroplating metalizes the surface of a target conductive object with a metal of choice depending on the application. For 3-D printed objects, the surface can be made conductive by spraying a base seed layer of nickel or silver conductive spray paint to allow the copper to build from. MG Chemicals Super Shield Nickel Conductive Coating [20] was used as the base metallizer. Copper plating was then performed to the 3-D printed ground planes.

The part to be electroplated was submerged in a copper sulfate acid solution, with copper sheets surrounding the part. These copper sheets are the anode source material that will be plated onto the target objects' surface at the cathode. To connect a wire directly to the surface of the plated part, MG Chemicals Silver Conductive Epoxy 8331 [21] was used to electrically connect a wire to the part that can be plugged into the power supply. The sulfate solution strips copper ions from the copper plates submerged on the perimeter, and the electrical current attracts them to the target part, which is grounded. A diagram of an example electroplating setup can be seen in Figure 3.3.

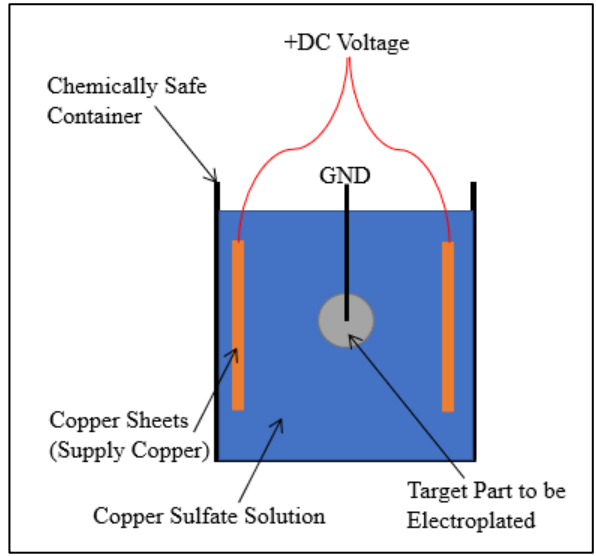

**Figure 3.3:** Diagram of electroplating setup for 3-D printed parts.

The formula for the copper sulfate solution used to electroplate is listed in Table 3-1. The brightener acts as a plating accelerator which can aid the deposition of the copper layer and improve the smoothness of the surface. The power supply utilized for the electroplating process was the E3631A HP Triple Output DC Power Supply.

**Table 3-1:** Copper Sulfate formula for electroplating.

| Ingredient | Amount |
|---|---|
| Distilled Water | 1000 ml |
| Copper Sulfate | 240 g |
| Sulfuric Acid | 32 ml |
| Brightener | 3 ml |

# Chapter 4

## Antenna and Ground Plane Designs

### 4.1.1 Monopole Antenna Design Theory

A widely used, cheap design of a simple wire monopole antenna is of quarter-wavelength size mounted on a ground plane that is fed by a coaxial line [6]. The antenna element length is set to a quarter the resonant frequency wavelength. Figure 4.1 depicts a diagram of a quarter wavelength monopole antenna mounted vertically and perpendicular from a flat ground plane.

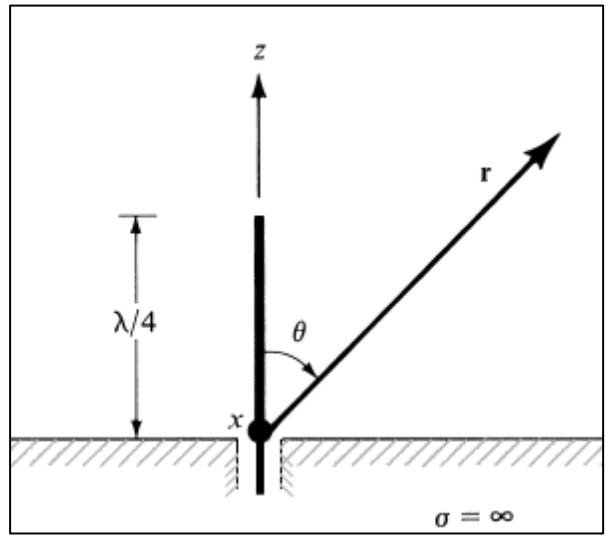

**Figure 4.1:** Quarter wavelength monopole antenna mounted on flat ground plane [6].

A RG402 coaxial cable was used where the antenna element was the inner conductor of the coaxial line. The outer conductor and inner insulation, Teflon, was stripped away to precisely a quarter of the resonant frequency wavelength. Below the antenna element where the outer conductor was still intact on the coaxial line, is where the ground plane encounters the antenna system. Figure 4.2 depicts the cross-sectional profile of a coaxial line displaying the inner conductor and insulating Teflon. Note this is before the coaxial line was mounted to a proper ground plane.

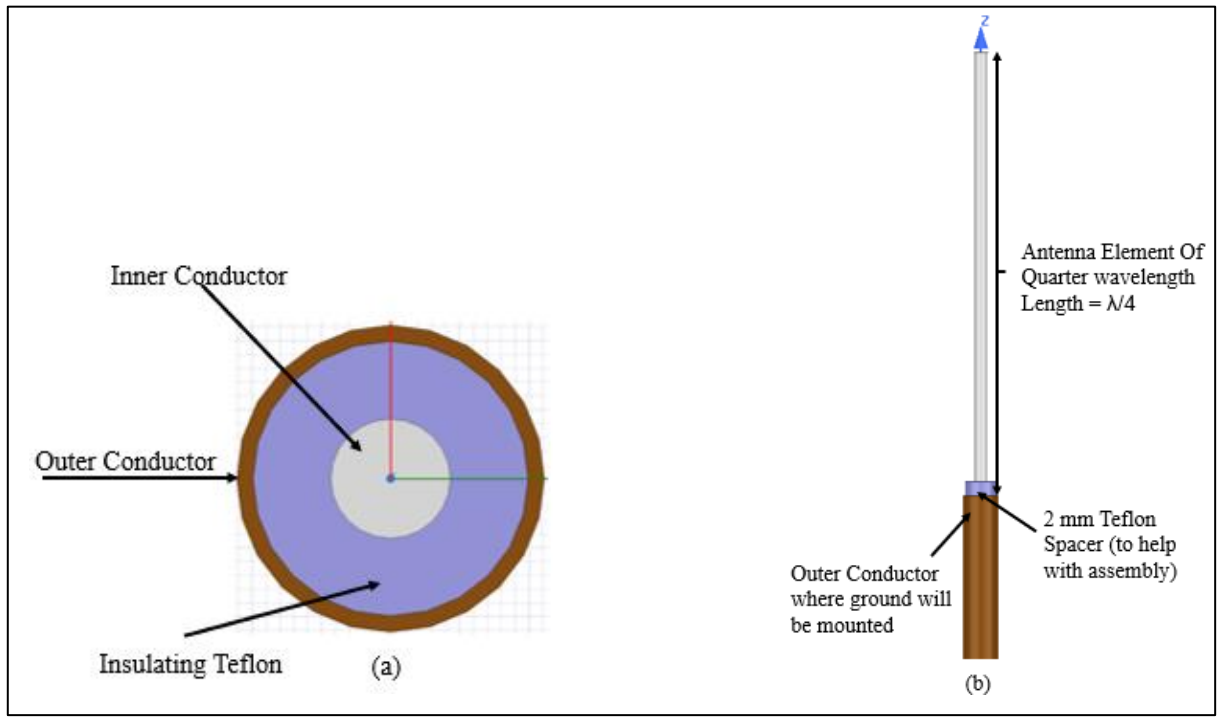

**Figure 4.2:** (a) Cross section of coaxial line, and (b) coaxial line with exposed inner conductor.

There is a 2mm Teflon spacer that was not stripped shown on the coaxial line between the antenna element and the outer conductor, that was left to aid with the assembly process. The Teflon buffer is to prevent a short from occurring when the ground plane is epoxied to the outer conductor. The antenna was designed to operate at 1.3GHz, with a wavelength of 230mm, where the antenna element was be a quarter of the wavelength, or 57.5mm.

## 4.1.2 Modeling of the Monopole Antenna

Simulations of the antenna's performance as well as experimental simulations for the multiple altered ground plane designs have been conducted in a full wave simulator. The model for the coaxial line was constructed by measuring the diameters of the inner Teflon and conductor and making solid cylinders matching in size. The dimensions of the coaxial wire antenna that was used in the simulations are illustrated in Figure 4.3. When a ground plane is mounted to the outer coaxial conductor, it is placed flush with the outer conductor where it meets the 2mm buffer of Teflon.

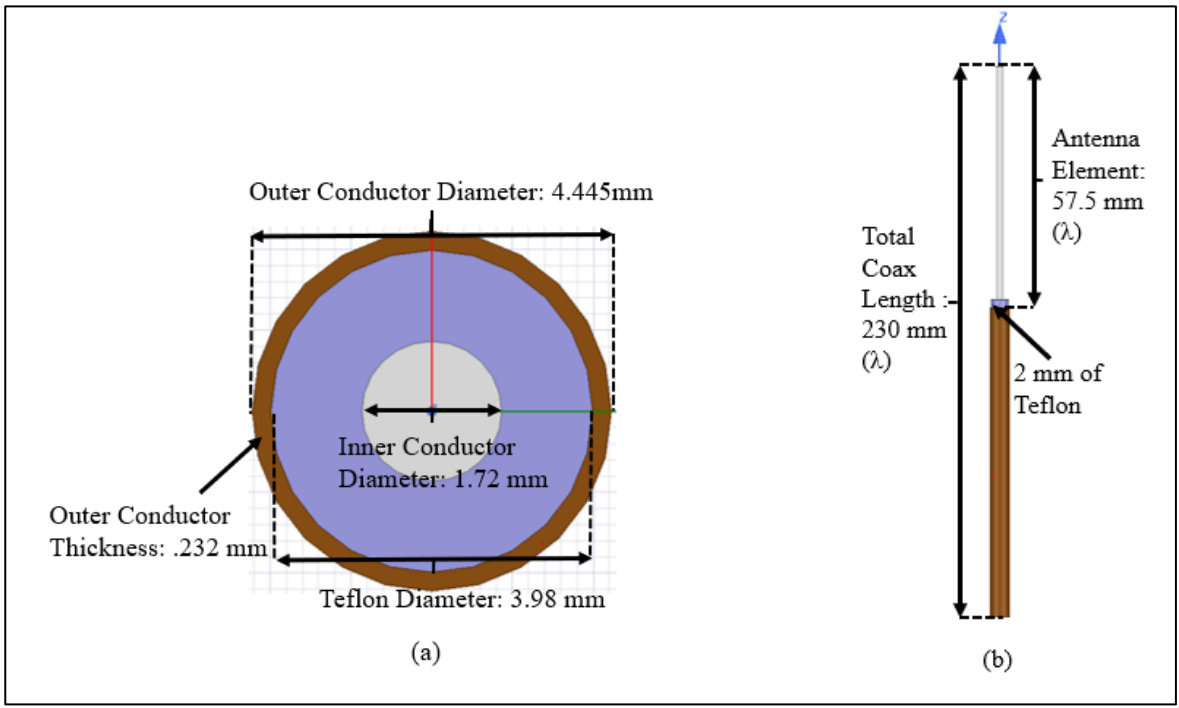

**Figure 4.3:** (a) Cross section dimensions of coaxial line, and (b) coaxial line with dimensions.

A RG402 male SMA connector was used for the feed which can be seen in Figure 4.4. The length of copper conductor to be stripped for the male SMA connector pin was about 0.2mm. This allows the male SMA pin to slip onto the inner conductor prior to soldering, allowing a reliable connection. Once the male pin is soldered on, the remaining threaded piece of the connector can be pressed onto the coaxial cable. To ensure a good physical and electrical connection, MG Chemicals Silver Conductive Epoxy 8331 [21] is applied where the connector and coax outer conductor meet.

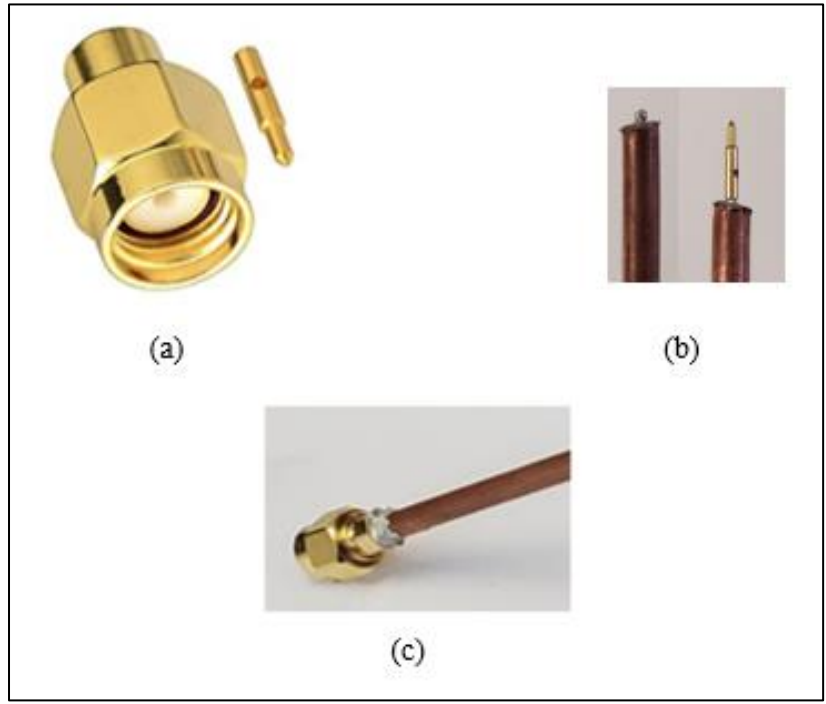

**Figure 4.4:** (a) RG402 Male SMA connector, (b) prepared coax and mounted SMA pin on coax line, (c) SMA connector mounted to coax line.

After stripping away the inner conductor for the antenna element and affixing the male SMA connector to the other end of the coaxial line, the ground plane is ready to be mounted, completing the fabrication process. The ground plane is secured to the coaxial line with the MG Chemicals Silver Conductive Epoxy 8331 [21] which ensures a stable physical and electrical connection.

4.1.4 Standard Ground Geometry Dimensions

  The flat ground plane was used as one of the base reference geometries. The other ground plane shape examined was a spherical ground plane. Both ground plane geometries will have a minimum radius of a quarter wavelength, identical to the antenna element length. Each ground plane is 1.5mm in thickness. Figure 4.5 shows a top and isometric view of the base planar and spherical ground planes. Note that there is a mounting hole in the model of the radius to fit snugly around the outer conductor of the coaxial line.

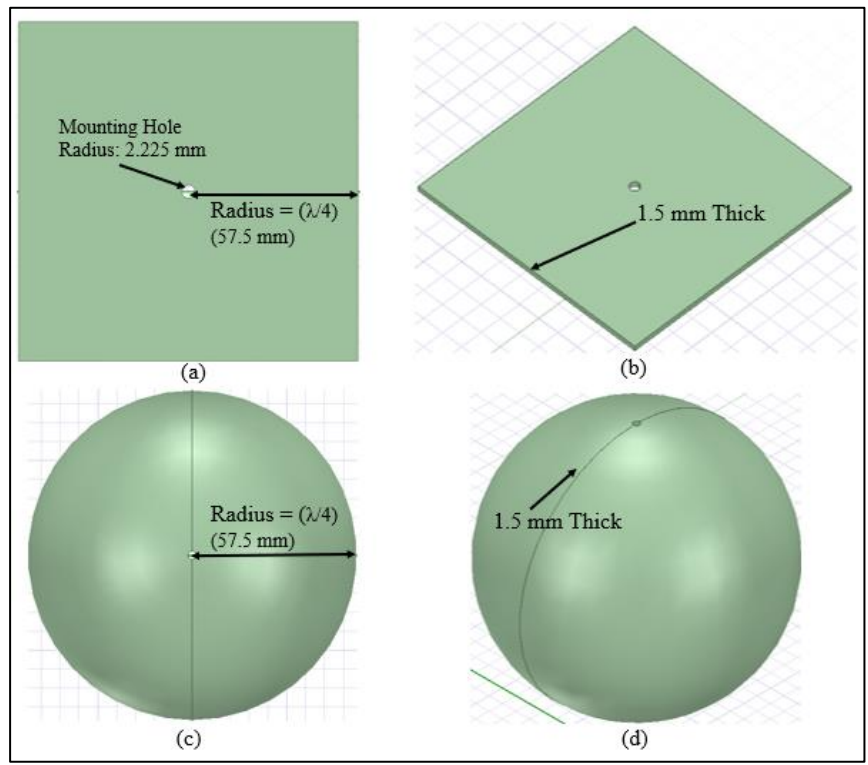

**Figure 4.5:** Dimensions of base ground planes: (a) planar top view, (b) planar isometric view, (c) sphere top view, (d) sphere isometric view.

### 4.1.5 Standard Ground Geometry Modeling

The fabrication of the planar ground plane was a copper sheet of appropriate thickness cut to size, and the antenna mounting hole was drilled in the center. Copper spherical ground planes on the other hand were 3-D printed using ABS material and were metallized. Initially, metallization of the surface was done using 3 coats of the MG Chemicals 843AR Silver-Coated Copper Conductive paint [22]. However, upon measurement, gain measured was significantly deteriorated illustrating an insignificant thickness for metallization coat. Therefore, the 3-D printed ground plane structures were initially painted on the surface of the spheres post printing, then electroplated with copper to add more metal thickness. The skin depth of copper at 1.3GHz is only 1.8E-3mm, therefore, to ensure the copper layer is thick enough, the default thickness of the copper layer for simulation was chosen to be 0.1mm.

## 4.2 Monopole Simulation

### 4.2.1 Ideal Planar Ground

The first model simulated is a monopole antenna mounted to an infinitely large, perfectly conducting flat planar ground. The purpose of this simulation is to verify the constructed model will produce the expected theorical E radiation pattern in EM simulation. The simulated E radiation pattern can be seen in Figure 4.6 with the theorical pattern for comparison. It is evident that the generated E-plane pattern from the simulation is in good agreement with the theoretical pattern. The two circular main lobes are on either side of the polar plot that are symmetrical. Judging from the E radiation pattern found in the simulated ideal planar ground on a coaxial monopole antenna, the simulation is set up properly and is producing expected radiation results.

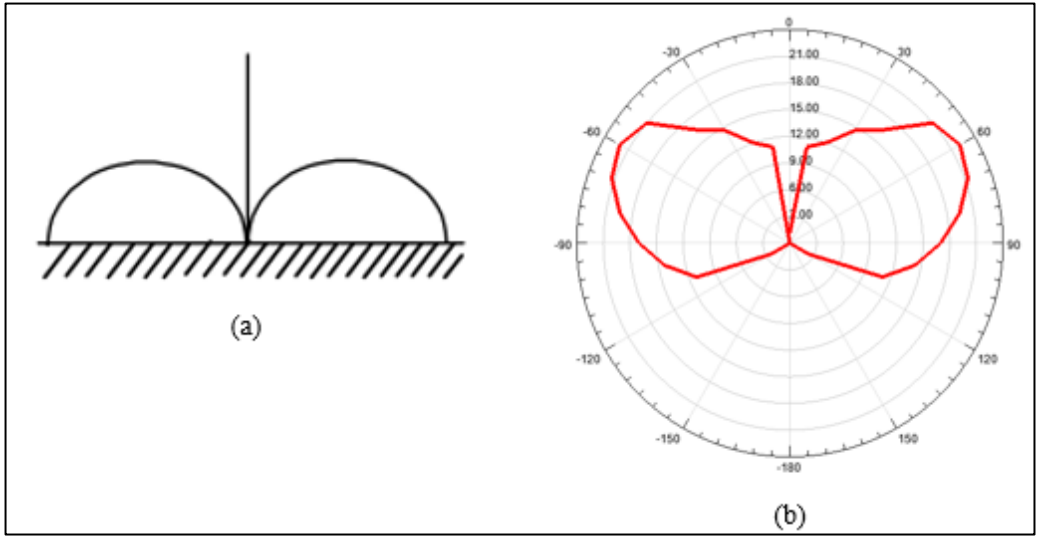

**Figure 4.6:** (a)Theoretical quarter wave planar E-Field radiation pattern [14], (b) simulated monopole above perfect conductor.

### 4.2.2 Standard Planar Ground

A simple flat copper planar ground plane with the dimensions of 115mm x 115mm (standard radius 57.5mm or ¼ $\lambda$) is shown in Figure 4.7 with the ground plane fixed to the coaxial line. The results of this simulation are given in Table 4-1, and Figure 4.8 displays the resonant frequency, -10dB bandwidth, antenna input impedance, gain and E-plane radiation pattern generated. Figure 4.9 shows the current distribution of the standard planar ground plane with the monopole antenna.

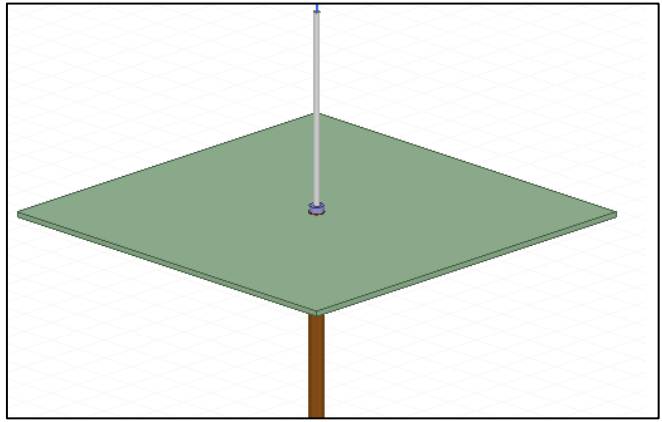

**Figure 4.7:** Simulation model of coaxial line antenna with copper planar ground mounted.

**Table 4-1:** Standard planar simulation results.

| Ground Plane Type | $S_{11}$ at Resonant Frequency (GHz) | $Z_{in}$ ($\Omega$) | -10dB Bandwidth (%) | Gain (dB) |
|---|---|---|---|---|
| Standard Planar | -11dB at 1.3 | 45.2+26.3j | 6 | 3.8 |

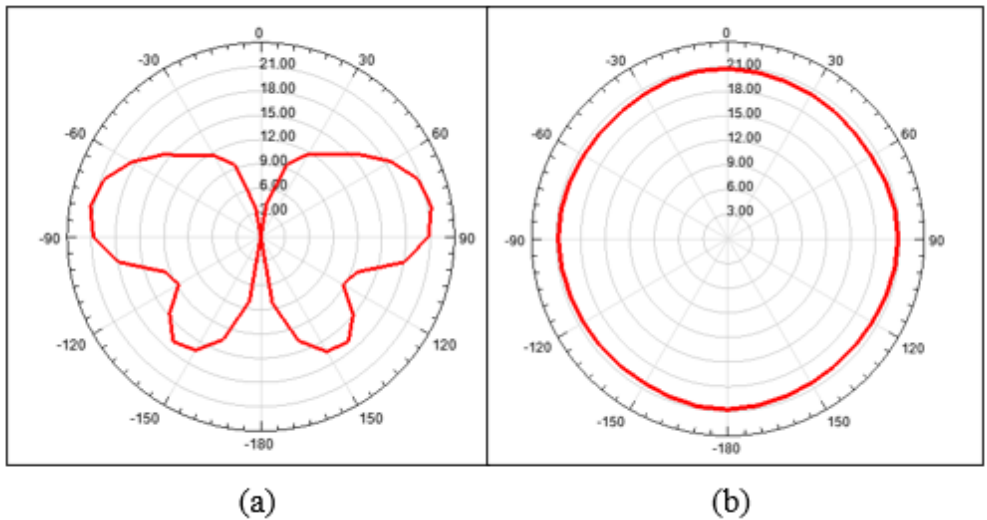

**Figure 4.8:** Simulated E-Pattern(a) and H-Pattern(b) of standard planar ground.

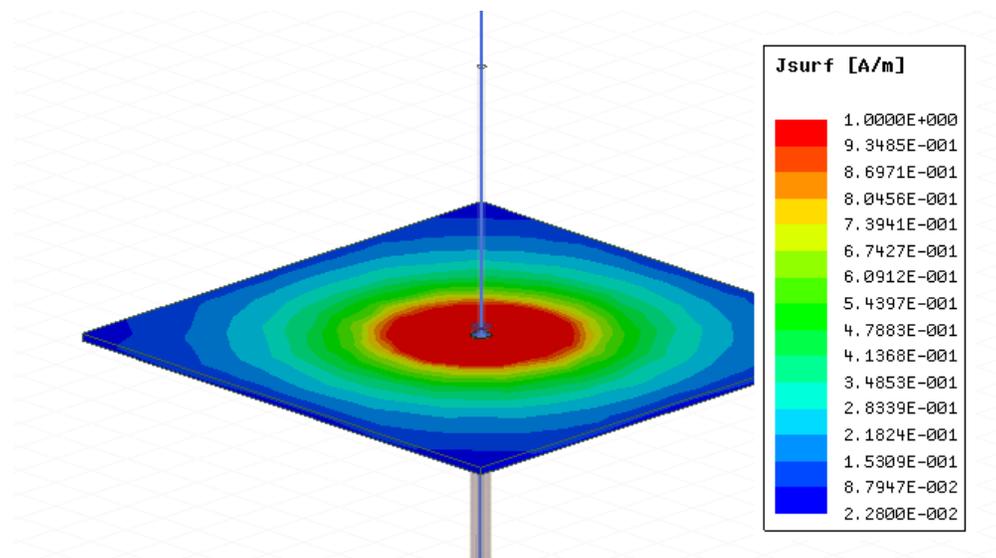

**Figure 4.9:** Standard planar current distribution.

The monopole had a resonant frequency of 1.3GHz, which was the target frequency in the design. The gain was found to be 3.8dB, while the -10dB bandwidth was 6%. The current distribution in Figure 4.9 shows the current dispersing evenly in a circular pattern away from the antenna. The E radiation pattern generated is like that of the ideal plane, but with addition of minor back lobes. This may be due to the finite ground plane radiating some energy from the induced currents [16].

### 4.2.3 Standard Spherical Ground

Figure 4.10 shows an isometric view of the standard quarter wavelength sphere mounted to the coaxial monopole antenna. The results are listed in Table 4-2, and the E Plane radiation pattern is shown in Figure 4.11.

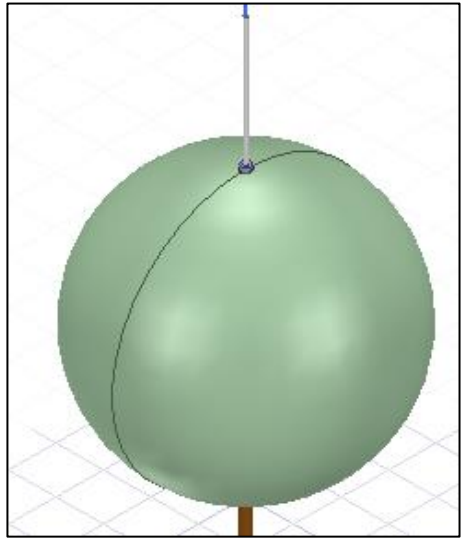

**Figure 4.10:** Model of coaxial line antenna with standard spherical ground plane for EM simulation.

**Table 4-2:** Standard spherical ground plane simulation results.

| Ground Plane Type | $S_{11}$ at Resonant Frequency (GHz) | $Z_{in}$ ($\Omega$) | -10dB Bandwidth (%) | Gain (dB) |
|---|---|---|---|---|
| Standard Sphere | -15dB at 1.3 | 46.7-15.8j | 22 | 3.4 |

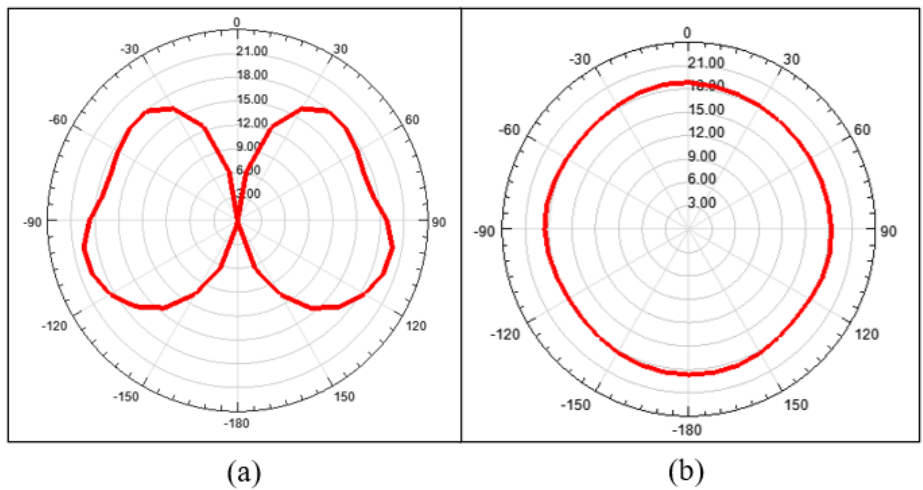

**Figure 4.11:** Simulated standard spherical E-Pattern(a) and H-Pattern(b) radiation pattern.

The bandwidth of the standard sphere was four times as large as the standard planar at 22% bandwidth. The simulated gain was 3.4dB. The E-field radiation pattern has 2 main lobes only that are stretched towards the top of the polar plot compared to the simulated standard planar having 2 main lobes and 2 minor lobes reaching towards the front. Furthermore, the planar $Z_{in}$ has a more inductive imaginary component while the spherical $Z_{in}$ has a capacitive imaginary component. By observing the current distribution along the surface of the sphere shown in Figure 4.12, most of the current is focused on the top of the sphere, closest to where the antenna is mounted. This is where alteration will be made to the ground plane in hopes of disrupting the current flow.

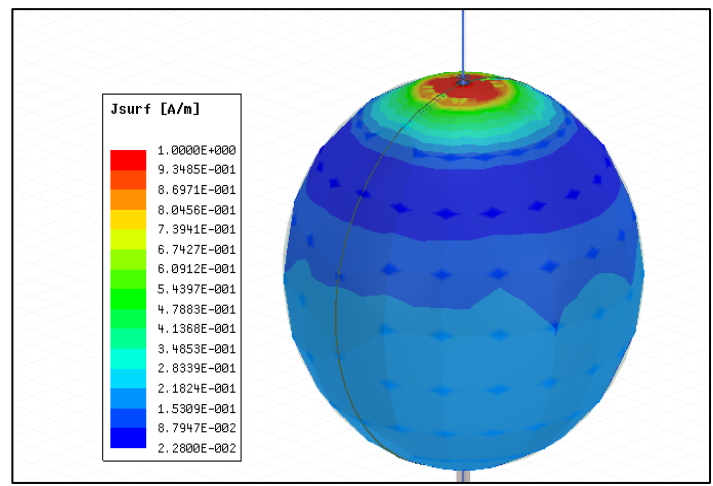

**Figure 4.12:** Surface current of spherical ground plane from EM simulation.

4.3 Altered Ground Planes

Using the base geometry of the standard planar and standard sphere shown previously, alterations were made to try to enhance the monopole antenna's performance. The same coaxial antenna model was used for each design, and only the ground planes are modified in each simulation. If the radiation pattern is not shown for an altered ground plane, the reason is that it was the same as the standard ground plane.

### 4.3.1 Ribbed Planar

The ribbed planar ground plane is like the standard planar, with the addition of spiked ribs protruding along the surface. A profile, top and isometric view of the ribbed planar ground plane can be seen in Figure 4.13 along with the dimensions of the spiked ribs. The motivation of this design was to mimic the potential location of a heat sink near the antenna and its impact on the antenna's performance. The ribs are a 1/64 of the operating wavelength in height. They are quite narrow in width, 1mm. The findings after simulating the model can be seen in Table 4-3 along with the standard planar simulation results.

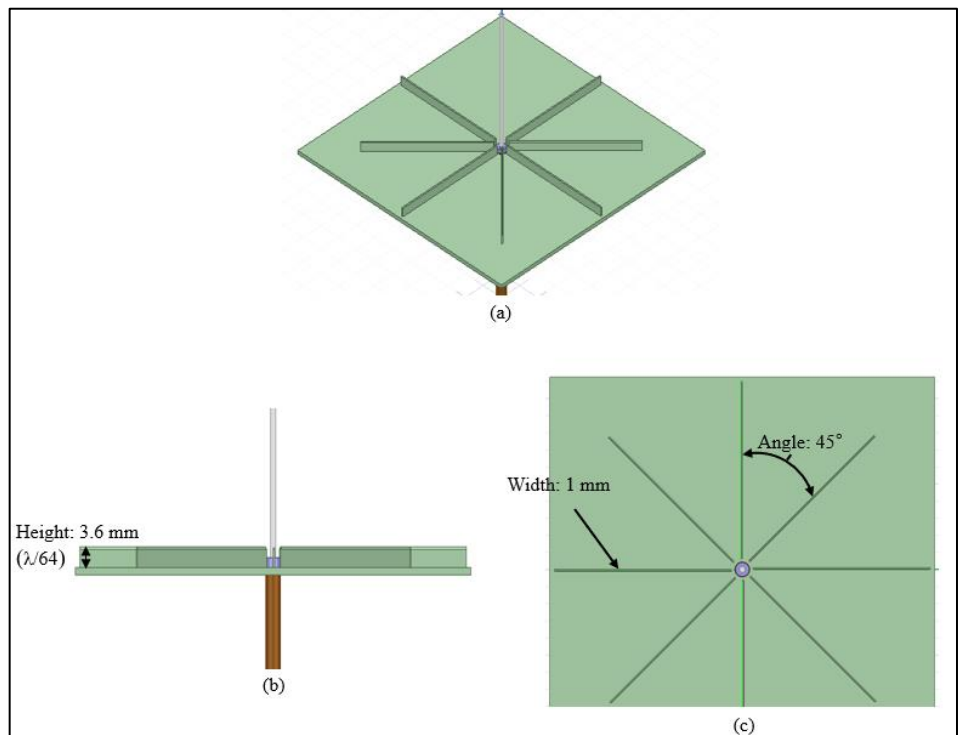

**Figure: 4.13:** (a) Isometric, (b) side, (c) top view of the ribbed planar ground plane with labeled dimensions.

**Table 4-3:** Ribbed planar and standard planar simulation results.

| Ground Plane Type | $S_{11}$ at Resonant Frequency (GHz) | $Z_{in}$ ($\Omega$) | -10dB Bandwidth (%) | Gain (dB) |
|---|---|---|---|---|
| Ribbed Planar | -14dB at 1.3 | 66+17.5j | 8 | 4 |
| Standard Planar | -11dB at 1.3 | 45.2+26.3ij | 6 | 3.8 |

The ribbed planar ground did not significantly improve the characteristics of the antenna compared to the standard planar. There was a slight increase in bandwidth, resonant frequency, and the return loss improved by 3dB. The gain, and E-field radiation pattern of the Ribbed planar showed no change compared to that of the standard planar ground plane.

4.3.2 Planar with Dish

The next modified ground plane that was simulated was a planar ground with a dish as shown in Figure 4.14 with isometric, side, and top view of the ground plane. The purpose of the parabolic dish is to reflect parts of the radiated and received signals back to the antenna element in hopes of focusing the signal so that it becomes stronger. The flat ground plane footprint was maintained at standard size of 115mm by 115mm.

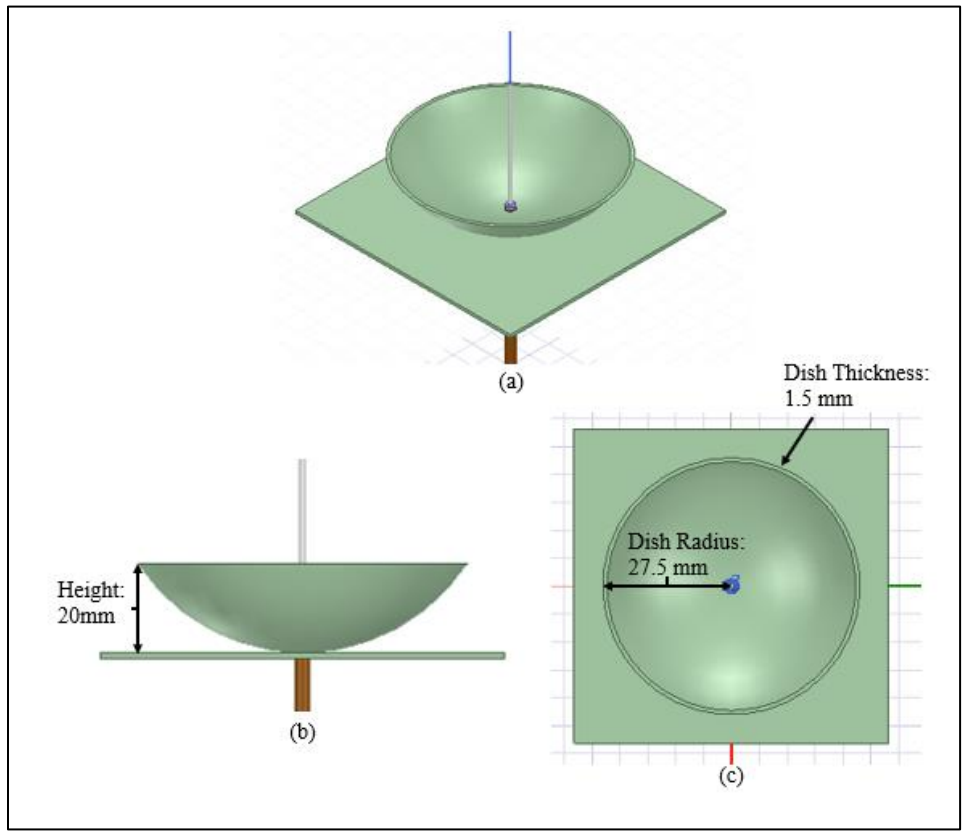

**Figure 4.14:** (a) Isometric, (b) side, and (c) top view of planar dish ground.

To determine the height and radius of dish, optimization was performed. The height and radius of the dish were varied from 20mm to 55mm in steps of 5mm. Table 4-4 depicts parametric sweep results of the changes in diameter and Table 4-5 depicts the sweep of height. Figure 4.15 depicts the changes in E field radiation pattern with change in height. Radius and height were tested independently, so when the diameter was under test, the height was fixed at the chosen value of 20mm, and vice versa. Then once all optimized dimensions are realized, a higher resolution simulation is performed with only the optimized dimensions applied.

**Table 4-4:** Results of sweeping radius of dish in EM simulation (height set to 20mm).

| Diameter of Dish (mm) | $Z_{in}$ (Ω) | Bandwidth (%) | Gain (dB) |
|---|---|---|---|
| Planar (NO DISH) | 45.2+26.3j | 6.5 | 3.8 |
| 20 | 49.6+17j | 11.9 | 3.7 |
| 25 | 49.7+15.7j | 12.3 | 3.7 |
| 30 | 49.7+14.66j | 13 | 3.7 |
| 35 | 49.3+13.6j | 13.4 | 3.9 |
| 40 | 49.5+12.6j | 14.2 | 4 |
| 45 | 48+11.5j | 15 | 4.0 |
| 50 | 48.2+10.75j | 15 | 4.1 |
| 55 | 47.5+9.5j | 15.3 | 4.1 |

**Table 4-5:** Results of sweeping height of dish in EM simulation (diameter set to 20mm).

| Height of Dish (mm) | $Z_{in}$ (Ω) | Bandwidth (%) | Gain (dB) |
|---|---|---|---|
| Planar (NO DISH) | 45.2+26.3j | 6.5 | 3.8 |
| 20 | 37.8+1j | 17.6 | 4.2 |
| 25 | 23.1+.46j | 10 | 4.2 |
| 30 | 12+7j | N/A | 4 |
| 35 | 15+1+6j | N/A | 4 |
| 40 | 21.7+4j | N/A | 3.7 |
| 45 | 27+3j | N/A | 3 |
| 50 | 30.5+2.7j | N/A | 2.9 |
| 55 | 33.14+2.5j | N/A | 2.8 |

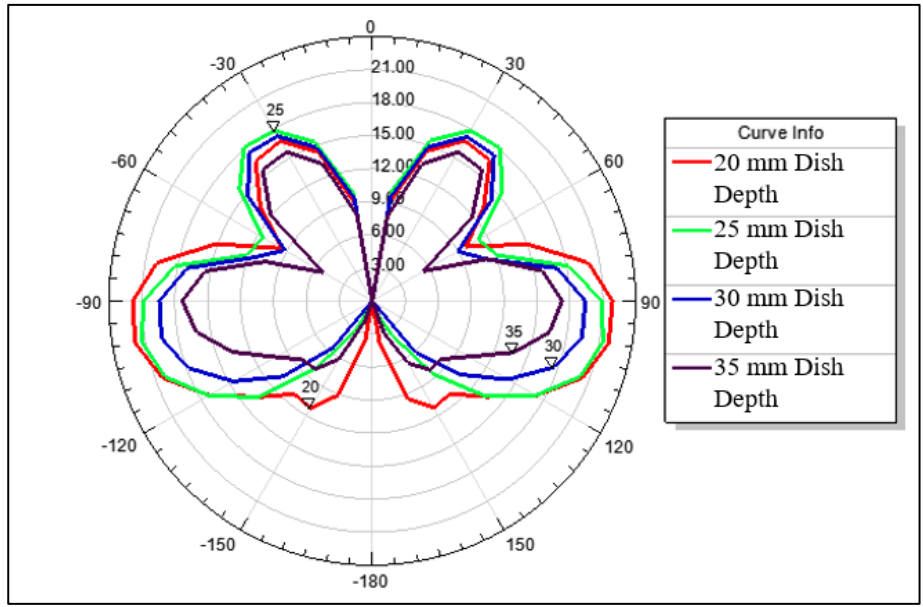

**Figure 4.15:** Radiation patterns of different planar dish height.

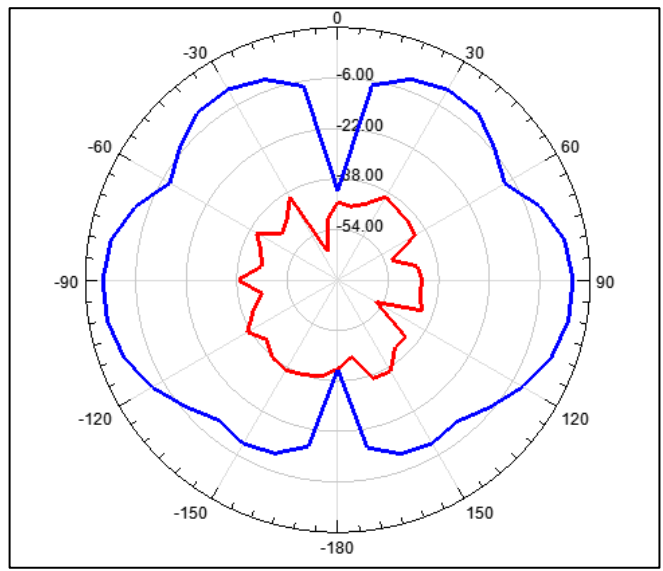

**Figure 4.16:** Co-polarization (Blue) and cross-polarization of planar dish (Red).

The inductive imaginary portion of the antenna's input impedance decreases as the radius increases. Furthermore, the bandwidth and gain reach a peak of 15% and 4.1dB, respectively. In terms of radius, the best option is 55mm in terms of bandwidth and gain performance. To realize the optimized dimensions of 55mm radius with height of 20mm, a sphere of 55mm radius is generated and dissected to remove a portion of the top half leaving a dish of 20mm height, as shown in a profile view in Figure 4.16. The blue half-sphere is of radius 55mm. To maintain an identical curvature to a sphere of radius 55mm while setting an ideal height of 20mm, the red portion is removed, and the final radius of the dish with a height of 20mm results in being 41.3mm.

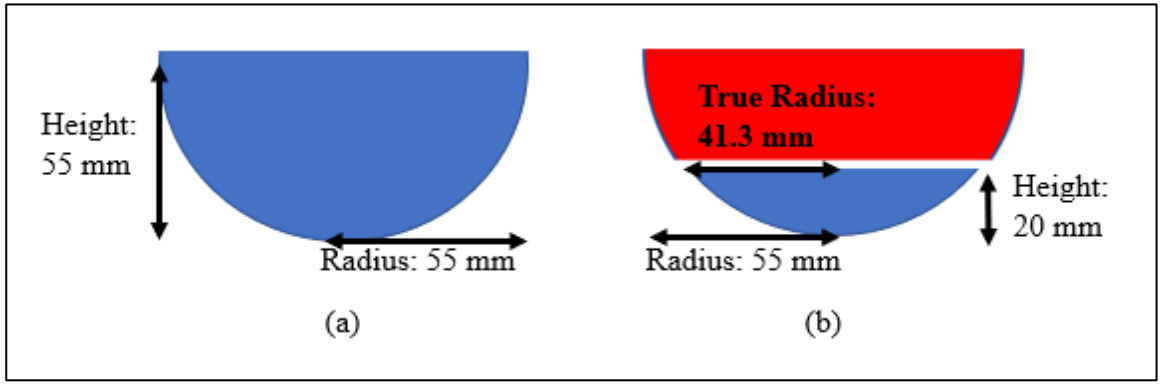

**Figure 4.17:** Illustration of how dish height affects true radius of the dish.

Table 4-5 depicts the results when sweeping different values for the height of the dish, where the optimal height is 20mm. After 25mm, the $S_{11}$ return loss performance deteriorates to less than -10dB. Furthermore, beyond the 20mm height, the gain diminishes as well. Figure 4.15 shows the radiation patterns as the height of the dish changes. The non-optimal radius radiation patterns have the same general shape as the standard planar, except the lobes are narrower. The intensity of the radiation pattern grows as the height reaches 20mm, where it peaks. Once the best dimensions were known, a detailed simulation was performed with the optimized dimensions of height 20mm and radius of 55mm (true radius 41.3mm). The results can be seen in Table 4-6 comparison to the standard planar. The planar dish with the optimized dimensions reaches a bandwidth of 20% in simulation which is almost four times higher than the standard planar. Furthermore, the return loss reaches -28dB compared to -11dB from the standard which is a drastic

increase in power performance. The gain slightly decreased to 3.5dB in comparison to the standard planar. The cross and co polarization plots can be seen in Figure 4.16. This design was also simulated with only the dish present (no copper plane), and it was found that the antenna didn't resonant past -3dB.

**Table 4-6:** Planar with dish and standard planar simulation results.

| Ground Plane Type | $S_{11}$ at Resonant Frequency (GHz) | $Z_{in}$ ($\Omega$) | -10dB Bandwidth (%) | Gain (dB) |
|---|---|---|---|---|
| Planar with Dish | -28dB at 1.3 | 40+1j | 20 | 3.5 |
| Standard Planar | -11dB at 1.3 | 45.2+26.3j | 6 | 3.8 |

### 4.3.3 Planar with Horn

With the performance improvement seen with the planar dish design, an additional geometry considered was the horn, known for its focusing and highly directive properties. The horn takes the form of a square flaring waveguide. An isometric, side and top view can be seen in Figure 4.18 along with its dimensions. The upper length and width of the horn was set to twice the lower length to produce the angled flat horn shape. Furthermore, through simulation it was found that the horn performs best in the form of square, where the width and length are the same value, instead of a rectangle. The length was varied from 15mm to 35mm in steps of 5mm. Tables 4-7 and 4-8 show the parametric sweep results found for the changing of the lower length and horn height respectively. The horn produced a compressed radiation pattern compared to that of a standard planar. The radiation patterns as the height changed is shown in Figure 4.19.

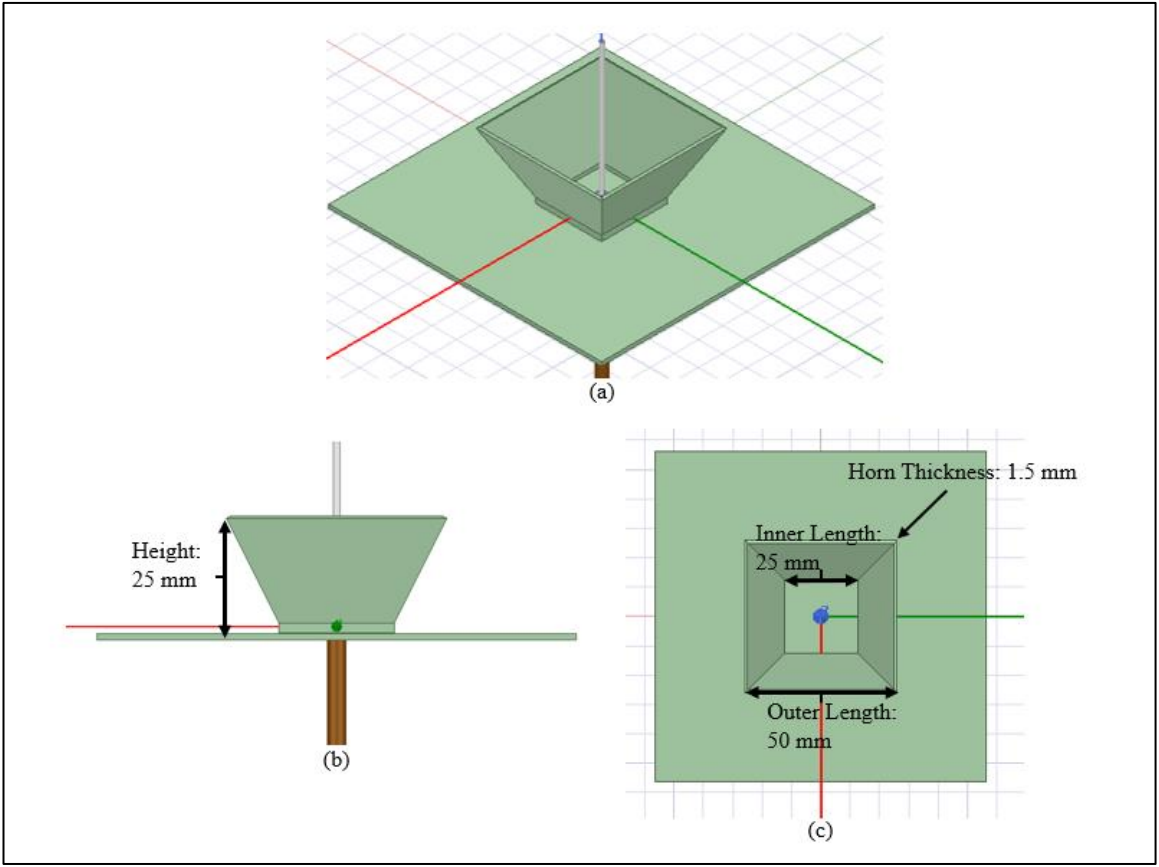

**Figure 4.18:** (a) Isometric, (b) side, and (c) top view of planar horn ground.

**Table 4-7:** Results of sweeping lower length of horn in EM Simulation (height set to 20mm).

| Lower Length of Horn (mm) | $Z_{in}$ (Ω) | Bandwidth (%) | Gain (dB) |
|---|---|---|---|
| Planar (NO DISH) | 45.2+26.3j | 6.5 | 3.8 |
| 15 | 39.6-.3j | 13.4 | 2 |
| 20 | 36+1.7j | 15 | 2.4 |
| 25 | 32.35+4.8j | 17.7 | 4.4 |
| 30 | 29.1+8.9j | 14.6 | 2.1 |
| 35 | 26.7+14j | 10.3 | 1.2 |

**Table 4-8:** Results of sweeping height of horn in EM Simulation (length set to 20mm).

| Lower Length of Horn (mm) | $Z_{in}$ (Ω) | Bandwidth (%) | Gain (dB) |
|---|---|---|---|
| Planar (NO DISH) | 45.2+26.3j | 6.5 | 3.8 |
| 15 | 43+16j | 13.4 | 1.7 |
| 20 | 37.3+12.2j | 15.6 | 2.2 |
| 25 | 29.12+9j | 20 | 3.2 |
| 30 | 20+8j | 14.6 | 2.4 |
| 35 | 10.3+10j | 10.7 | 1.9 |

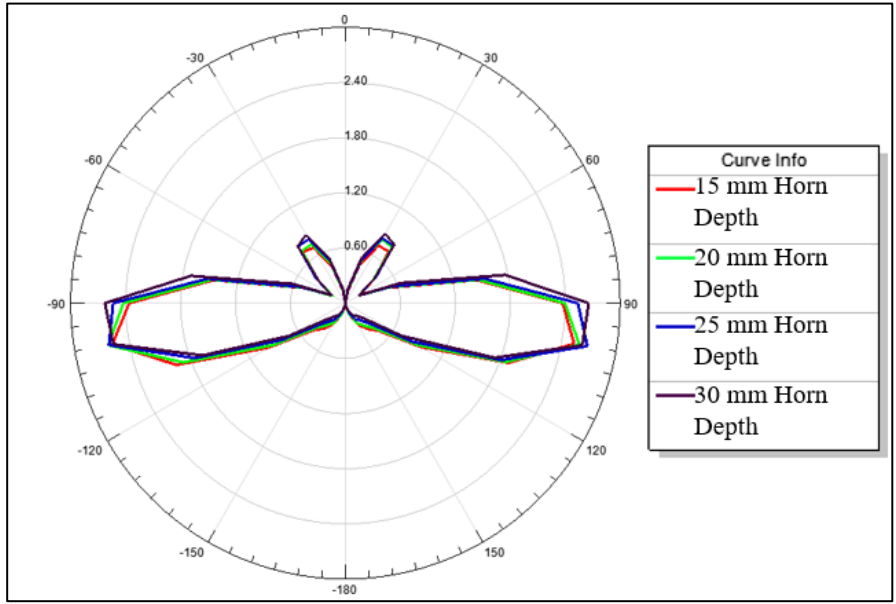

**Figure 4.19:** Radiation patterns of different planar horn height.

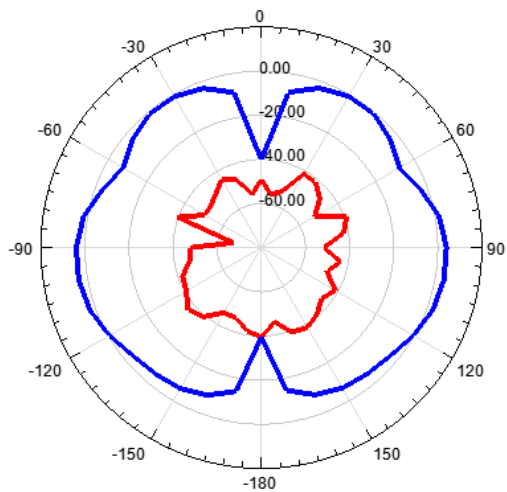

**Figure 4.20:** Co-polarization (Blue) and cross-polarization of planar with horn.

The peak bandwidth and gain across the swept parameters of lower length and horn height occurred at 25mm. The optimized dimension results for the horn antenna can be seen in Table 4-9 compared to the standard planar. The gain diminished slightly, as well as the real component of $Z_{in}$. However, the imaginary component is much closer to 0, and the return loss greatly improved to -31dB instead of -11dB. The radiation pattern produced, which is shown in Figure 4.19, is narrowed, and stretched towards -90° and 90° on the polar plot. It is difficult to see each individual radiation pattern since they are quite similar. The smaller plots are heights of 15mm while the largest plot is the 25mm peak. The cross and co polarization plots can be seen in Figure 4.20. This design was also simulated with only the horn present (no copper plane), and it was found that the antenna didn't resonant past -5dB.

**Table 4-9:** Planar with horn and standard planar simulation results.

| Ground Plane Type | $S_{11}$ at Resonant Frequency (GHz) | $Z_{in}$ ($\Omega$) | -10dB Bandwidth (%) | Gain (dB) |
|---|---|---|---|---|
| Planar with Horn | -31dB at 1.3 | 32.35+4.8j | 20 | 3.2 |
| Standard Planar | -11dB at 1.3 | 45.2+26.3j | 6 | 3.8 |

### 4.3.4 Planar with Cone

Another simple geometry that could exhibit reflective properties is a conical structure. It has similar properties to the dish and horn in that it looks as if it could reflect waves from the inside of the cone back towards the antenna element, hopefully increasing signal strength and intensity. A diagram showing an isometric, top and side view of the conical ground plane can be seen in Figure 4.21 along with the dimensions of the cone. Parametric optimization was performed on the radius and height of the cone. The height and radius were varied from 20mm to 45mm in steps of 5mm, independently. Tables 4-10 and 4-11 show the parametric sweep results found for the changing of the cone radius height respectively. Figure 4.22 shows the simulated E radiation pattern.

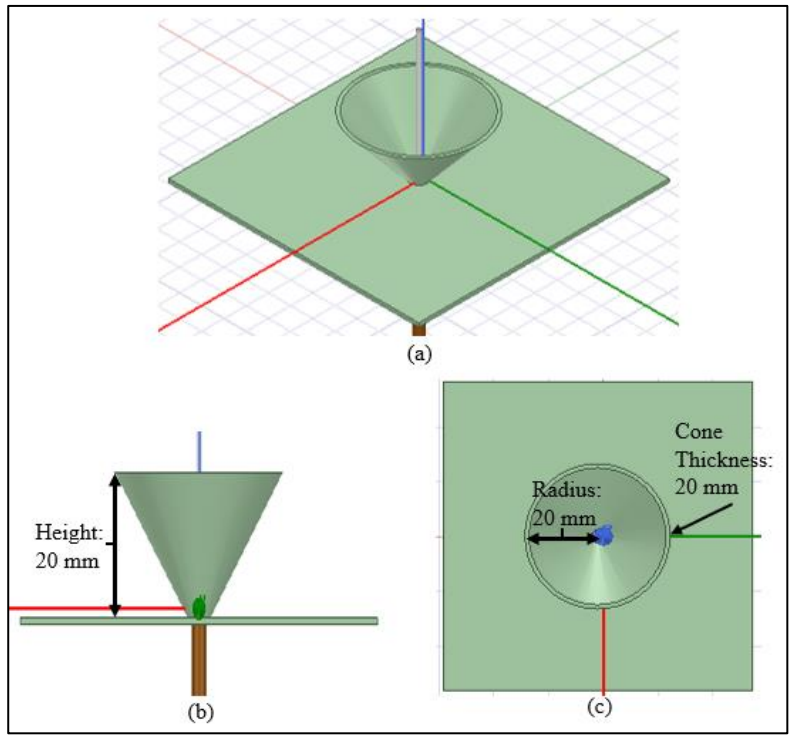

**Figure 4.21:** (a) Isometric, (b) side and (c) top view of planar with cone ground.

**Table 4-10:** Results of sweeping radius of cone in EM simulation (Height set to 20mm).

| Radius of cone (mm) | $Z_{in}$ ($\Omega$) | Bandwidth (%) | Gain (dB) |
|---|---|---|---|
| Planar (NO DISH) | 45.2+26.3j | 6.5 | 3.8 |
| 15 | 52.7-12j | 2.5 | 1.4 |
| 20 | 56.3-10i | 2.2 | 1.3 |
| 25 | 55.9-11.3j | 2.5 | 1.3 |
| 30 | 59-15j | 2.8 | 1.2 |
| 35 | 54.2-14j | 2.7 | 1.2 |
| 40 | 60.7-18j | 2.8 | 1.1 |
| 45 | 62.1-16.4j | 2.2 | 1 |

**Table 4-11:** Results of sweeping radius of cone in EM simulation (Radius set to 20mm).

| Height of cone (mm) | $Z_{in}$ ($\Omega$) | Bandwidth (%) | Gain (dB) |
|---|---|---|---|
| Planar (NO DISH) | 45.2+26.3j | 6.5 | 3.8 |
| 15 | 54.6-12j | 2.8 | 1.2 |
| 20 | 51.3-10i | 2.3 | 1.2 |
| 25 | 57-11.3j | 2.5 | 1.1 |
| 30 | 55.9-15j | 2.2 | 1.0 |
| 35 | 54-14j | 2.2 | 1.0 |
| 40 | 52.8-18j | 1.5 | .7 |
| 45 | 58.4-16.4j | 1.2 | .76 |

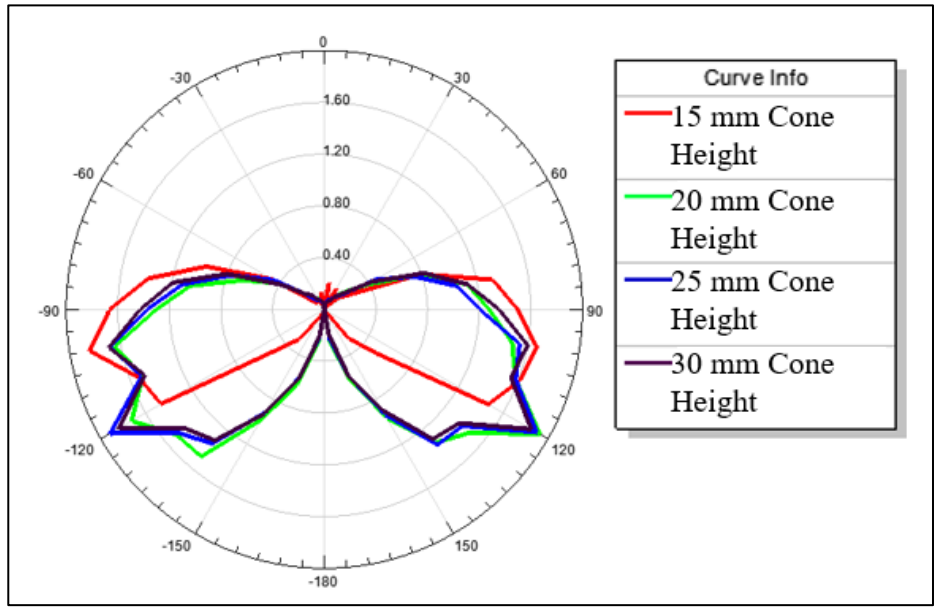

**Figure 4.22:** Radiation patterns of different planar cone height.

It is clear from Tables 4-10 and 4-11 that the cone antenna didn't perform well in terms of bandwidth and gain. All the bandwidths were under 3% and the peak gain was 1.4dB. The interesting observation of the cone planar ground is that the resonant frequency shifted much higher; to 3.15GHz instead of 1.3GHz. This is a 1.85GHz shift upward in resonance. Furthermore, the E field plane radiation pattern took a narrower shape that is directed towards the rear of the polar plot as the cone height increases.

### 4.3.5 Size of Spherical Ground Plane

The first modification to the spherical ground planes examined were changes in the radius of the ground plane. Four different radii were simulated in addition to the standard sphere which was a quarter wavelength in radius. Table 4-12 lists radii of the spheres as a fraction of the wavelength as well as in millimeters. The results from the multiple sphere size simulations are listed in Table 4-13 which also includes the previous simulation of the standard sphere, which is quarter wavelength radius.

**Table 4-12:** Different sized sphere radii used in EM simulation.

| Sphere Radius ($\lambda$) | Sphere Radius (mm) |
|---|---|
| $1/8$ | 28.75 |
| $1/4$ | 57.5 |
| $3/8$ | 86.25 |
| $1/2$ | 115 |
| $3/4$ | 174.8 |
| 1 | 230 |

**Table 4-13:** Different sized sphere radii EM simulation results.

| Radius of Sphere | $Z_{in}$ ($\Omega$) | Bandwidth (%) | Gain (dB) |
|---|---|---|---|
| $1/8\lambda$ | 61-20j | 17 | 2.8 |
| $1/4\ \lambda$ | 46.7-15.8j | 14.6 | 3.44 |
| $3/8\ \lambda$ | 52.9-11.6j | 17 | 2.8 |
| $1/2\lambda$ | 37.7-10j | 15 | 2.55 |
| $3/4\lambda$ | 49.9-.9j | 15 | 2 |
| $1\lambda$ | 45.9+18j | 13 | 1.9 |

The simulation states that as the radius of the sphere increases, the gain and bandwidth diminish. Furthermore, the smaller sphere has a more capacitive impedance while the largest sphere has an inductive $Z_{in}$. However, as the radius increases past half a wavelength, the sphere's $S_{11}$ plot gains multiple resonance frequencies compared to the single resonance frequencies the standard quarter wavelength sphere has. Figure 4.23 displays the $S_{11}$ plots of the $3/4\lambda$ and $1\lambda$ spheres as the short and long dashed lines, respectively. The $3/4\lambda$ sphere has two clear deep resonant frequencies at both 1.17GHz and 1.25GHz. The full wavelength sphere has two shallower dips of resonance at 1.33GHz and 1.4GHz.

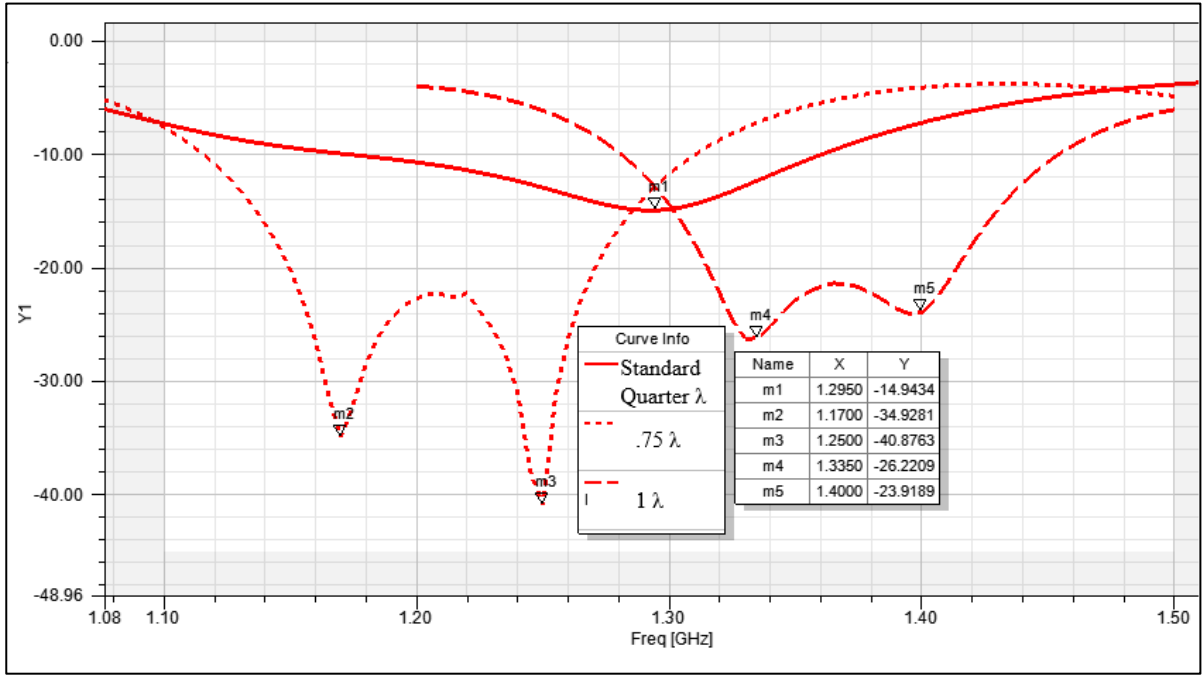

**Figure 4.23:** $S_{11}$ plot of larger diameter spheres to display multiple resonant frequencies.

### 4.3.6 Slotted Sphere

A method to attempt to redirect or disrupt current is by choosing where copper is deposited on the sphere during electroplating. As discussed previously, once the sphere is 3-D printed, it needs to be made conductive to electroplate. Conductive paint is applied. By carefully placing precisely cut pieces of masking tape along the surface of the sphere prior to spraying on the base seed conductor, there can be designs made along the surface that won't have copper deposited during electroplating. This leaves exposed ABS as the surface rather than a metalized surface. The first approach was simple long rectangular slots protruding away from antenna mount location. The slots are mounted close to the antenna element since this is where most of the current is concentrated. Figure 4.24 shows an isometric, side, and top view of the slotted spherical ground plane.

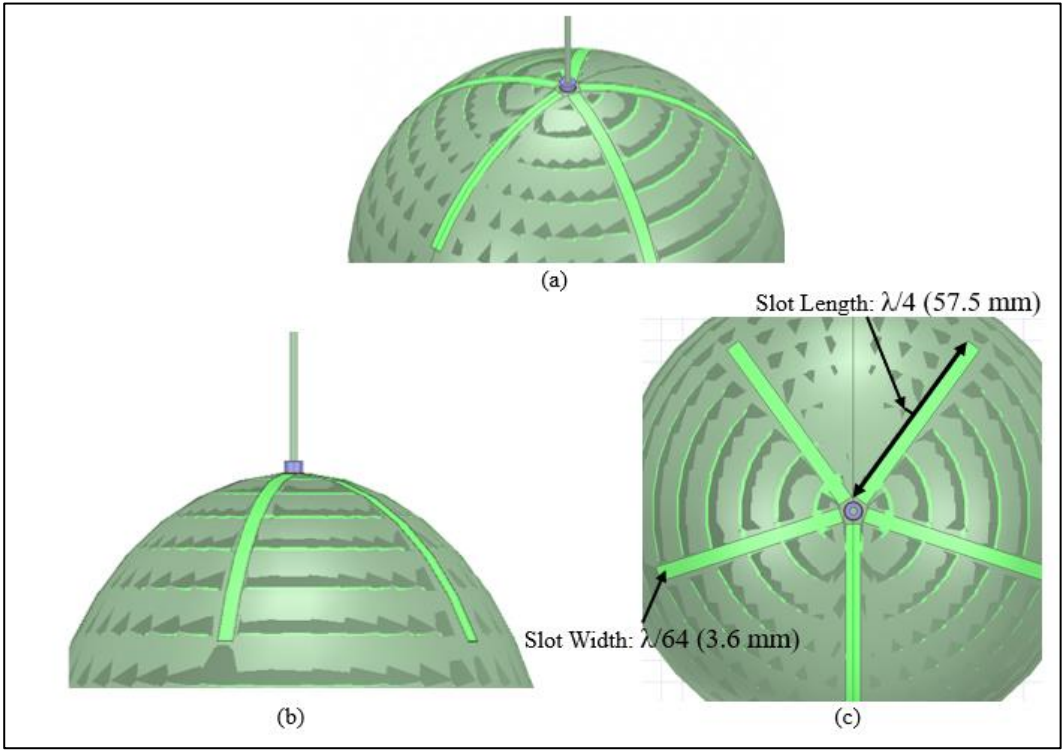

**Figure 4.24:** (a) Isometric, (b) side, and (c) top view of slotted spherical ground.

The results of the slotted quarter wave sphere in comparison to the standard quarter wave are seen below in Table 4-14. The resonance frequency dropped to 1.2GHz, while the return loss improved marginally. The gain had a slight decrease while the bandwidth had an increase in performance. The interesting result from the slotted quarter wave was it seems to eliminate half of the capacitive component on the impedance of the standard sphere. Furthermore, Figure 4.25 depicts the standard sphere surface current next to the slotted sphere surface current. The slots appeared to disrupt the current distribution as it avoided the ABS slots and causing an irregular distribution of the current compared to the standard sphere where it is consistent and smooth. The E radiation pattern remained unchanged compared to the standard sphere.

**Table 4-14:** Slotted sphere and standard sphere simulation results.

| Ground Plane Type | $S_{11}$ at Resonant Frequency (GHz) | $Z_{in}$ ($\Omega$) | -10dB Bandwidth (%) | Gain (dB) |
|---|---|---|---|---|
| Slotted Sphere | -19dB at 1.2 | 45.2-7j | 20 | 2.6 |
| Standard Sphere | -15dB at 1.3 | 46.7-15.8j | 14.6 | 3.4 |

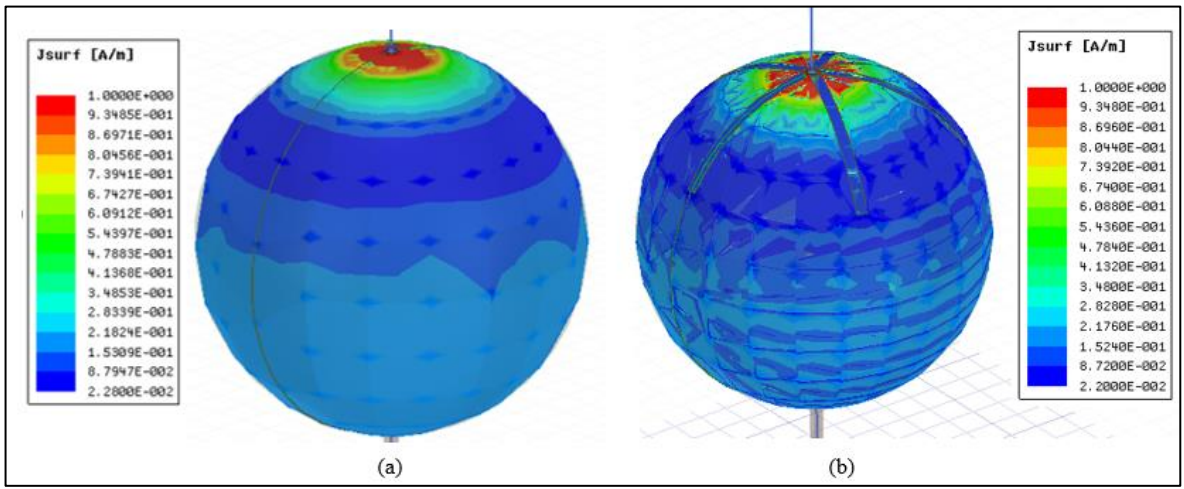

**Figure 4.25:** (a) Standard sphere surface current, (b) Slotted sphere surface current.

4.3.7 Ringed Sphere

A similar design to the slotted sphere in which copper is masked off from being plated on the ABS is the ring sphere. The ring sphere has horizontal rings of ABS that are exposed. Figure 4.26 shows an isometric, side, and top view of the slotted spherical ground plane along with dimensions of the rings and the distance between each ring. The intention was to again disrupt the current path by introducing ABS patches of varying width. The simulation results can be seen in Table 4-15.

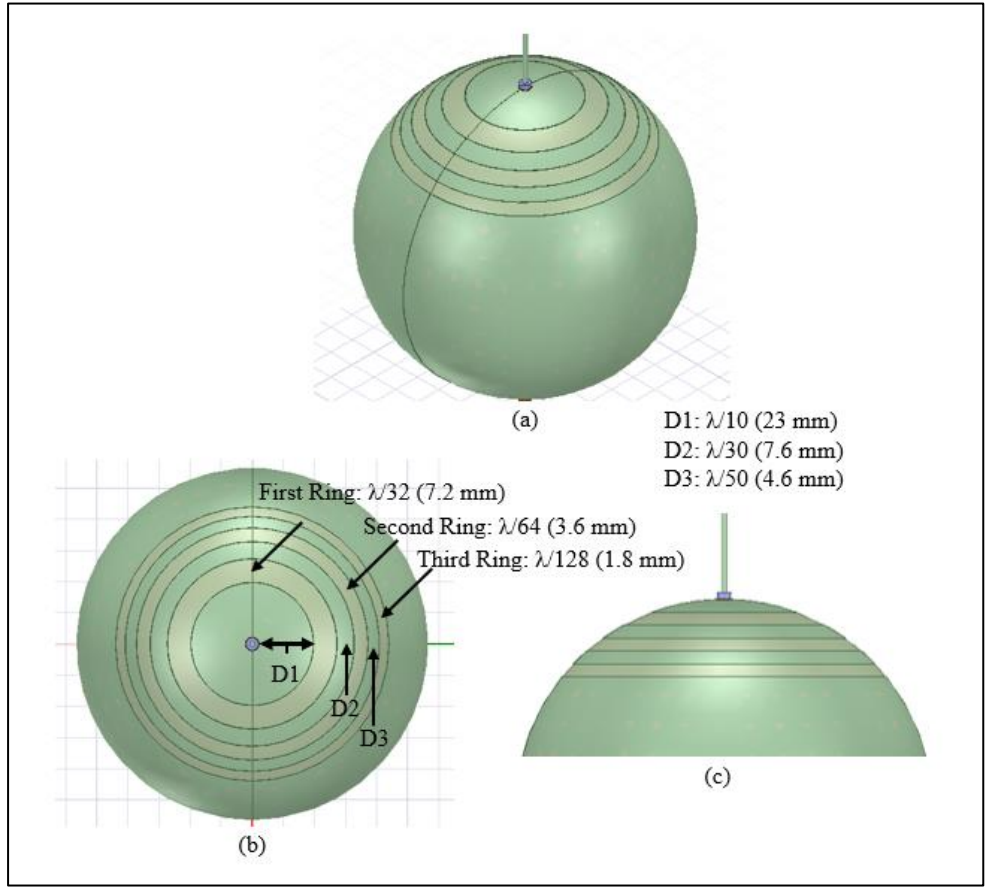

**Figure 4.26:** (a) Isometric, (b) side, and (c) top view of ringed spherical ground.

**Table 4-15:** Ringed sphere and standard sphere simulation results.

| Ground Plane Type | $S_{11}$ at Resonant Frequency (GHz) | $Z_{in}$ (Ω) | -10dB Bandwidth (%) | Gain (dB) |
|---|---|---|---|---|
| Slotted Sphere | -15dB at 1.2 | 38-8j | 11 | 2.9 |
| Standard Sphere | -15dB at 1.3 | 46.7-15.8j | 14.6 | 3.4 |

The ringed sphere seemed to have a similar impact on the impedance that the slotted sphere did, it eliminated half of the capacitive component of Zin. However, the match and bandwidth suffered, with no improvement to the return loss. The real part of the impedance fell to 38Ω while the bandwidth was also reduced to 11%. The gain dropped as well from 3.4 dB on the standard sphere to 2.9 dB on the ringed sphere. The E radiation pattern remained unchanged compared to the standard sphere. Figure 4.27 depicts the standard sphere surface current next to the ringed sphere surface current. The rings appear to have forced a greater density of current towards the bottom of the sphere. The dark blue intensity extends much further down the sphere than in the standard sphere current distribution.

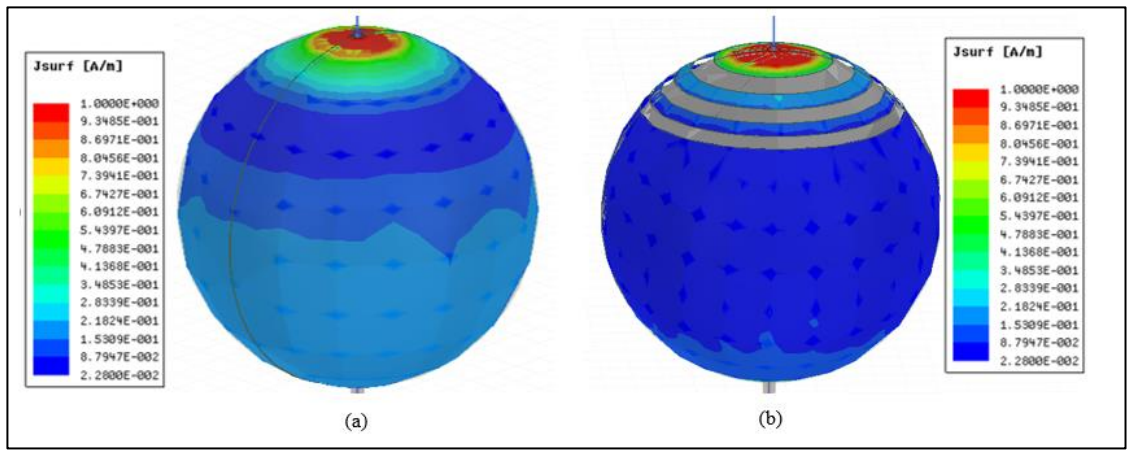

**Figure 4.27:** (a) Standard sphere surface current, (b) ringed sphere surface current.

4.3.8 Edge-mounted Spherical

The next type of spherical ground monopole antenna that was tested was the same geometry as the standard sphere, except the ground plane was mounted through an "edge" of the sphere, rather than directly through the center. This theoretically should force the current to have to flow in a different direction than the standard spherical because most of the ground plane is in a different location relative to the antenna than the standard sphere. An isometric, side and top view of the edge mounted ground plane is shown in Figure 4.26. The antenna is mounted $\lambda/64$ or 3.6mm from the edge of the sphere. The results can be seen in Table 4-16 along with the E radiation pattern shown in Figure 4.29 compared to the standard sphere E and H radiation patterns.

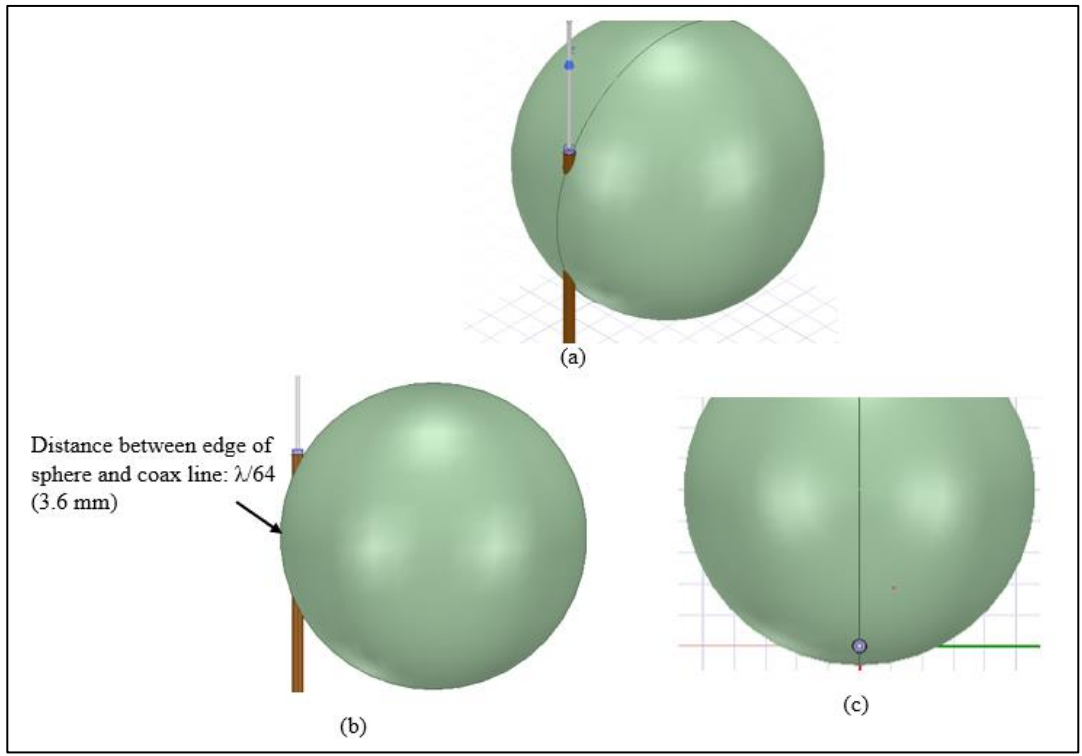

**Figure 4.28:** (a) Isometric, (b) side, and (c) top view of the edge-mounted spherical ground.

**Table 4-16:** Edge-mounted sphere and standard sphere simulation results.

| Ground Plane Type | $S_{11}$ at Resonant Frequency (GHz) | $Z_{in}$ ($\Omega$) | -10dB Bandwidth (%) | Gain (dB) |
|---|---|---|---|---|
| Edge-Mounted Sphere | -36dB @ 1.34 | 50.8+1.5j | 11 | 2.9 |
| Standard Sphere | -15dB @ 1.3 | 46.7-15.8j | 14.6 | 3.4 |

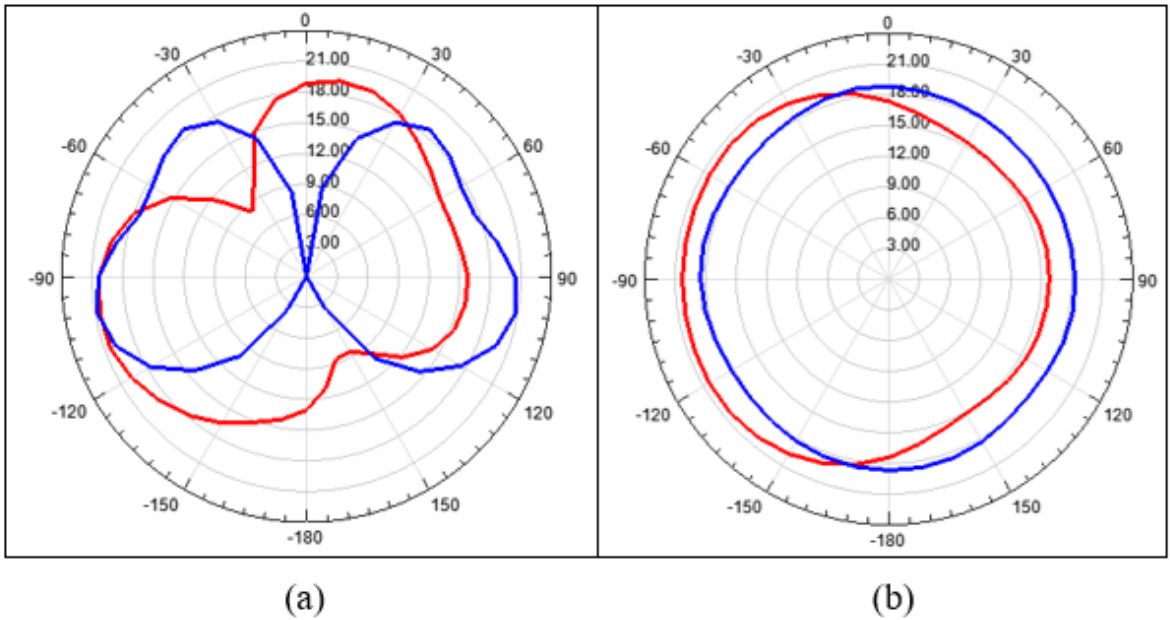

**Figure 4.29:** E plane (a) radiation pattern of edge mounted sphere (Red), compared to standard sphere (Blue), H plane radiation (b).

The edge mounted sphere has an impact on the return loss value. The return loss improved from -15dB in the standard sphere to -36dB when edge mounted. However, the bandwidth was reduced to 11% and the gain dropped slightly to 2.9dB. An interesting feature of the edge mounted location of the antenna is that it creates a directional radiation pattern that is directed towards where more of the ground plane is located, shown Figure 4.29. Therefore, by modifying where the antenna is mounted with respect to sphere ground, one can modify the radiation pattern of the antenna.

## 4.3.9 Sphere Composed of Fins

A rather abstract design of the standard spherical ground plane is having a polar array of short, quarter wavelength radius cylinders that are standing on their edge, rotated along the Z axis. For simple reference, these short cylinders will be referred to as fins. Figure 4.30 shows an isometric, side and top view of the sphere composed of fins along with the dimensions of the fins. The fin sphere was inspired from a similar design to a classic heatsink, maximizing the surface area within a given radius. The fins were chosen to be mounted vertically to be parallel to the antenna element. The results of the fin sphere can be seen in Table 4-17.

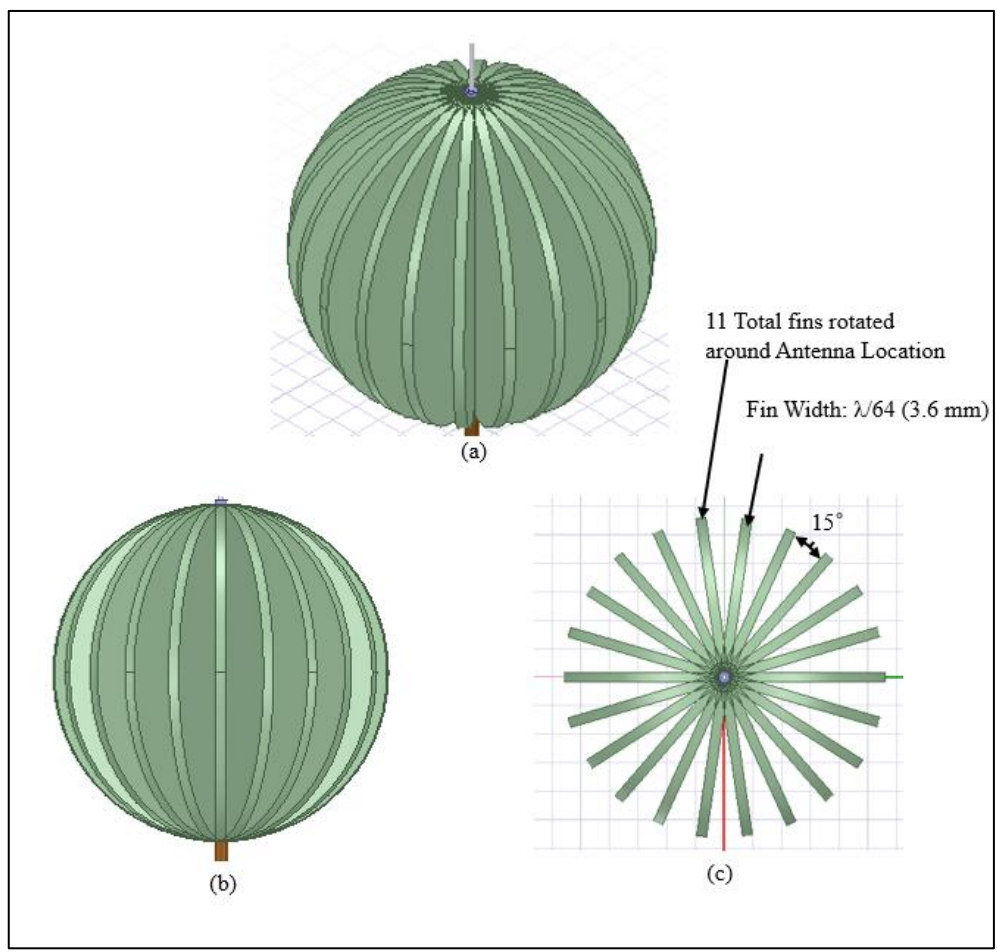

**Figure 4.30:** (a) Isometric, (b) side, and (c) top view of sphere composed of fins.

**Table 4-17:** Fin sphere and standard sphere simulation results.

| Ground Plane Type | $S_{11}$ at Resonant Frequency (GHz) | $Z_{in}$ ($\Omega$) | -10dB Bandwidth (%) | Gain (dB) |
|---|---|---|---|---|
| Fin Sphere | -18dB at 1.2 | 42-15j | 24 | 3 |
| Standard Sphere | -15dB at 1.3 | 46.7-15.8j | 14.6 | 3.4 |

It can be seen in Table 4-17 that the return loss slightly decreases while the bandwidth increases more dramatically in performance for the sphere constructed from fins compared to the standard sphere. The impedance remained the same while the gain dropped slightly. The radiation pattern remained unchanged compared to the standard sphere.

### 4.3.10 Spherical with Spikes

An additional alternative spherical design is adding additional small, thin cylinders that protrude from the surface of the standard ground sphere. The intention is to try to make the current flow up the thin spikes, changing its path than if it was going to travel on the surface of the sphere. Figure 4.31 shows an isometric, side and top view of the spiked sphere and its dimensions. The distances between rows and neighboring spikes were always fractions of a wavelength, as small as $\lambda/15$ and as large as $\lambda/4$. Figure 4.32 shows that the spikes did attract current and altered the wave nodes as the current traversed along the surface. However, all the spiked simulations performed identically to the standard sphere and had no impact on the performance of the sphere.

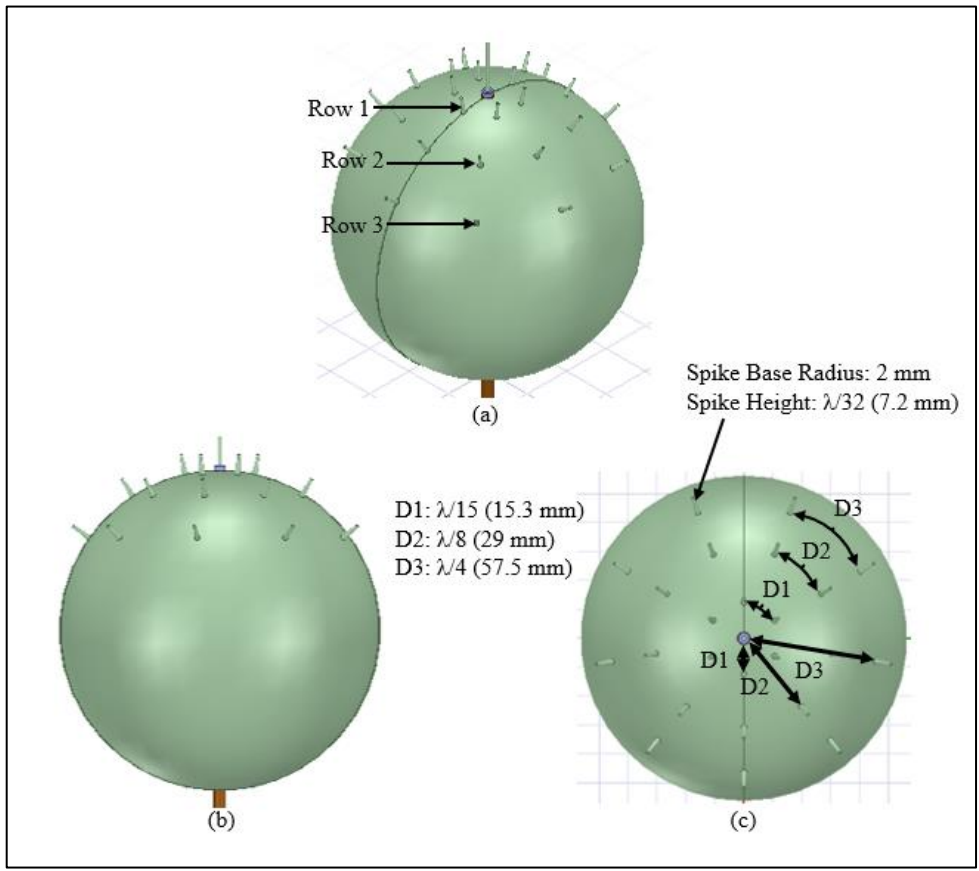

**Figure 4.31:** (a) Isometric, (b) side, and (c) top view of the spiked sphere.

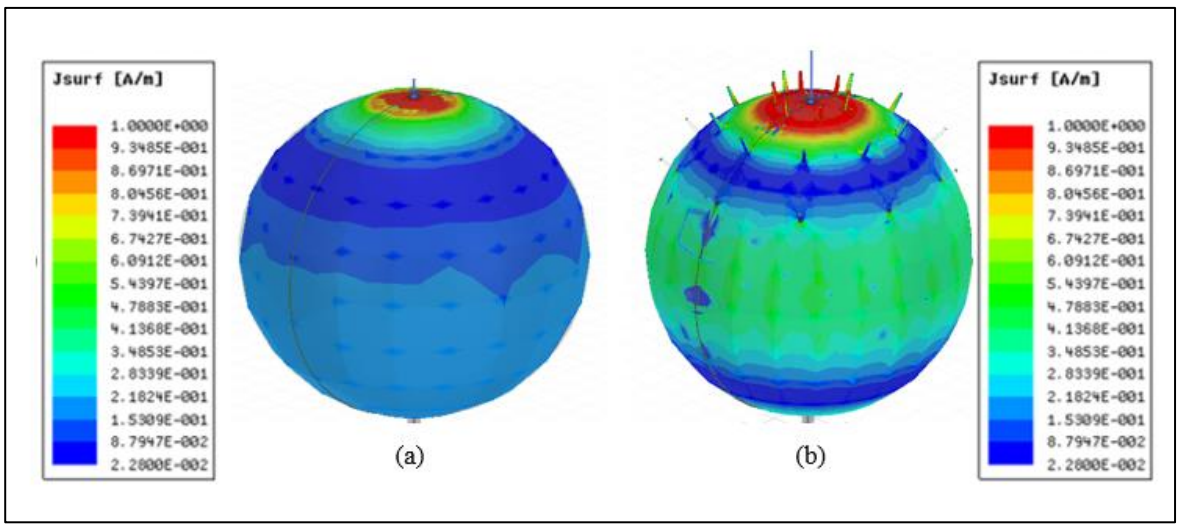

**Figure 4.32:** (a) Standard sphere surface current, (b) spiked sphere surface current.

# Chapter 5

## 5.1 Fabricated Standard Ground Plane Results

The standard planar and standard sphere are the first ground planes that were fabricated and tested. Moreover, the designs from the EM simulations that showed significant improvement over the standard ground planes were fabricated.

## 5.1.1 Standard Planar

To fabricate the standard planar antenna, a simple 1.5mm-thick copper sheet was cut to the proper dimensions of the standard planar ground as discussed previously. The radius was 57.5 mm, which means each length of the standard planar needs to be 115mm or 11.5cm. The copper sheet was cut to proper size using a benchtop sheet metal shearing tool. The mounting hole was drilled in the center of the copper sheet using a 5/16" drill bit. A tight-fitting mounting hole is preferable for best metal contact, but it also ensured the ground plane stays perpendicular to the coaxial line. Once the ground plane was set into proper position and held in place, silver epoxy was applied evenly to the top and bottom of the ground plane and the outer conductor of the coaxial line. This ensured a strong physical and electrical connection for the ground plane to the coaxial line. Figure 5.1 shows the fabricated standard planar. The measured results of the fabricated standard planar can be seen in Table 5-1 along with the simulated results. Figure 5.2 shows the measured E radiation field overlaid with the simulated E radiation plot.

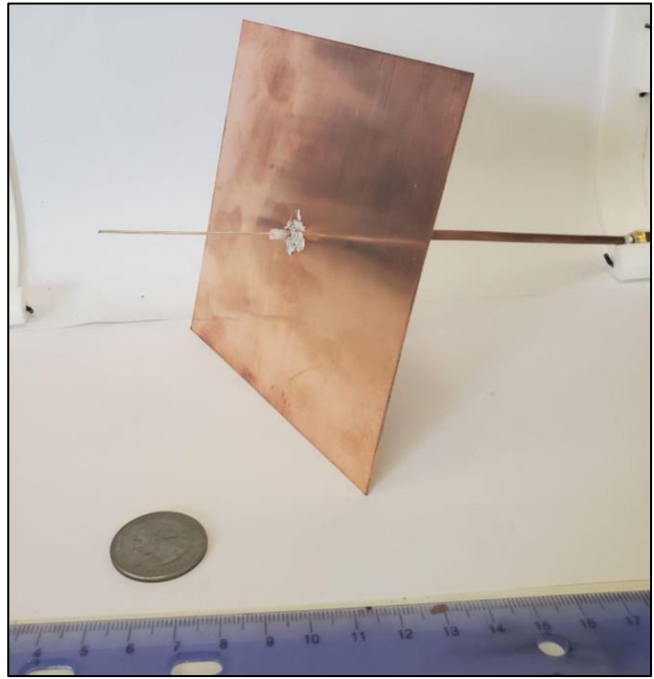

**Figure 5.1:** Fabricated standard planar.

**Table 5-1:** Measured results from fabricated standard planar with simulated.

| Ground Plane Type | $S_{11}$ at Resonant Frequency (GHz) | $Z_{in}$ ($\Omega$) | -10dB Bandwidth (%) | Gain (dB) |
|---|---|---|---|---|
| Measured Standard Planar | -10dB at 1.2 | 27.5-2.7j | 6 | 2 |
| Simulated Standard Planar | -11dB at 1.3 | 45.2+26.3j | 6 | 3.8 |

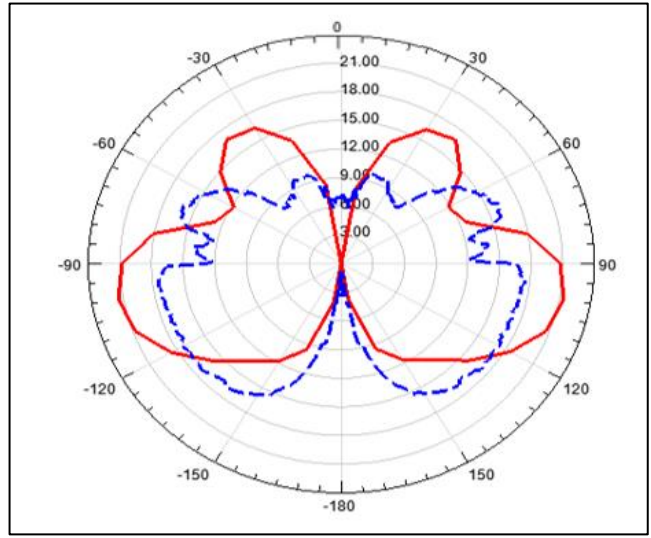

**Figure 5.2:** Measured E Field (Blue, Dash) and simulated (Red, Solid).

It can be seen from Table 5-1 that the fabricated standard planar performed quite similarly to the simulated. The measured return loss and bandwidth are almost identical to the simulated. The simulated planar had a better matched real Zin with an inductive imaginary component. However, the measured planar has a better matched imaginary component of -2.7j (which is close to 0), but a poorly matched real component of 27.5. The measured gain is half of what the simulation predicted. Furthermore, from observing Figure 5.2, the radiation pattern is in good agreement with the simulation as there are 2 main side lobes with 2 minor lobes in front of it. Differences in simulated and measured results may be a result of fabrication and assembly tolerances such as surface roughness and copper oxidation.

### 5.1.2 Standard Sphere Ground Plane Results

The construction of the sphere is more complex than the standard planar. The 3-D printed sphere needed to be acetone smoothed and nickel painted before it was ready to electroplate. To determine the optimal thickness of copper to be placed on the surface of the sphere, multiple identical spheres were electroplated with varying time and current intensity. Figure 5.3 displays one of the fabricated standard spheres. The spheres were then mounted on identical coaxial antennas and their RF characteristics were examined. The electroplating tests with the voltage, current and measured accumulated copper can be seen in Table 5-2. In the electroplating time column, if "Each Side " is seen, that means half of the part was electroplated at a time, that is the time per side. Otherwise, the entire part was submerged for the electroplating process. Table 5-3 shows the results of the electroplated spheres. The radiation pattern of Test Sphere 4 can be seen in Figure 5.4. The first test for metallization of the surface was MG Chemicals 843AR Silver-Coated Copper Conductive Paint, where three even coats were applied to the surface. The results of that sphere can also be seen in Table 5-3.

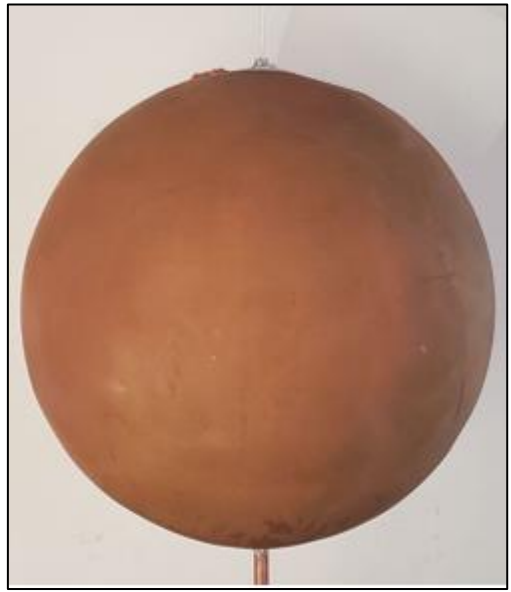

**Figure 5.3**: Fabricated standard sphere.

**Table 5-2:** Electroplating voltage, current, time, and accumulated thickness for standard sphere.

| Ground Plane Type | Electroplating Voltage (V) | Electroplating Current (A) | Electroplating Time (hrs) | Thickness of Copper (mm) |
|---|---|---|---|---|
| Test Sphere 1 | 5 | 1.2 | 1 (Each Side) | 0.3E-1 |
| Test Sphere 2 | 5 | 1 | 2 (Each Side) | 6.2E-2 |
| Test Sphere 3 | 5 | 1 | 4 | 0.1 |
| Test Sphere 4 | 5 | 1.2 | 4 | 1.3E-1 |

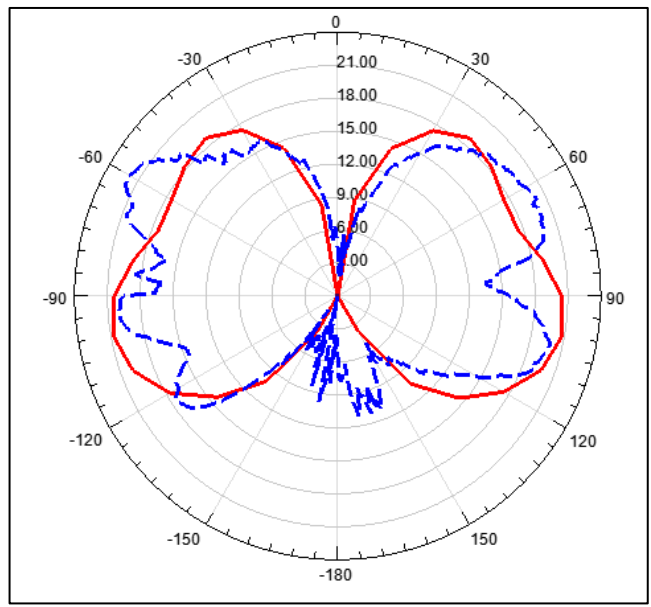

**Figure 5.4:** Measured E field from fabricated standard sphere (Blue, Dash) with simulated (Red, Solid).

It is observed from Table 5-2 that as the part electroplated for a longer time with a larger current, more copper was deposited. Table 5-3 shows that as more copper accumulates on the surface of the sphere, the antenna's RF performance improves. The first painted sphere and the first electroplated test sphere resonated; however, the gain and radiation pattern were immeasurable due to their low magnitude.

**Table 5-3:** Summarizing RF characteristics of electroplating tests on standard sphere.

| Ground Plane Type | $S_{11}$ at Resonant Frequency (GHz) | $Z_{in}$ ($\Omega$) | -10dB Bandwidth (%) | Gain (dB) |
|---|---|---|---|---|
| Simulated Standard Sphere | -15dB at 1.3 | 46.7-15.8j | 14.6 | 3.4 |
| Painted Sphere | -15dB at 1.3 | 49.5-1j | 12 | 0 |
| Test Sphere 1 | -25dB at 1.3 | 49.3-.2j | 15 | .2 |
| Test Sphere 2 | -27dB at 1.27 | 53.1-.5j | 16 | 1.7 |
| Test Sphere 3 | -35dB at 1.28 | 49.7+2.2j | 17 | 3.4 |
| Test Sphere 4 | -37dB at 1.28 | 50-2.4j | 16 | 3.3 |

Theoretically, the copper layer only needs to be as thick as the skin depth for copper at 1.3 GHz (1.8E-3mm). The first electroplated sphere had more copper than the skin depth, however it didn't perform nearly as well in terms of return loss and gain as when the sphere had .1mm of copper on its surface like test sphere 3 or 4. It seems that a relatively thick copper layer is needed over the ABS, otherwise the signal may be getting absorbed through the more abundant plastic base.

Once the plating thickness reached at least .1mm, the standard sphere RF characteristics performed reasonably well in comparison to the simulated. Even when the copper thickness went larger than .1mm, the performance didn't get much better compared to Test Sphere 3. For all other quarter wavelength spheres used in this thesis, a base current of 1.2Amps will be used along with 4 hours of electroplating time.

## 5.2 Fabricated Altered Ground Plane Results

### 5.2.1 Planar with Dish

To fabricate the planar dish, the same basic process was used, however the dish was fabricated separately and assembled on top of the planar ground plane. The dish was 3-D printed with the optimized dimensions found in simulation that was 55mm radius and 20mm in height of the dish. Since the dish was much smaller than the size of the standard sphere, the same electroplating settings couldn't be used because the standard sphere had a much larger surface area. However, the standard sphere electroplating time and current were used as a reference. The surface area of the standard sphere was found along with the surface area of the dish. The ratio of the two surface areas were used to determine the approximate amount of current to use for electroplating. After 2 hours of the electroplating process, the dish accumulated .11mm of copper on the surface. The dish was secured to the planar ground with the conductive silver epoxy. Figure 5.5 shows the fabricated planar dish antenna.

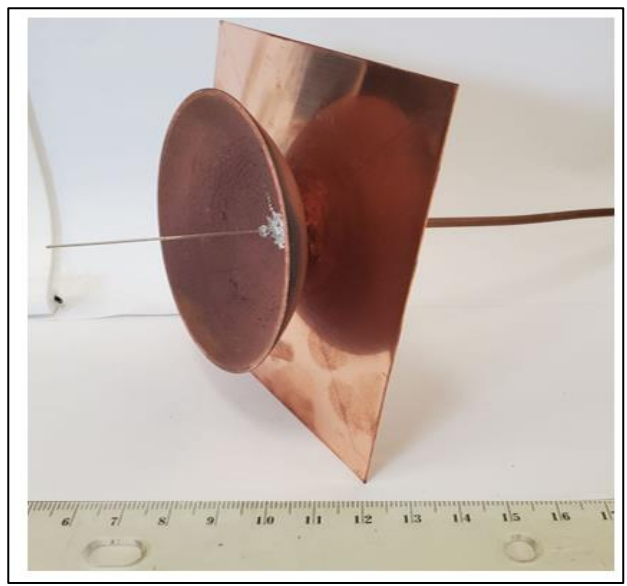

**Figure 5.5:** Fabricated planar with dish.

Table 5-4 depicts the measured results. Figure 5.6 shows the simulated radiation pattern compared to the measured. The bandwidth increased dramatically as predicted in the simulation to 24% rather than 6% from the standard planar. Furthermore, the match and return loss were greatly improved. The simulation predicted a large increase of -28dB over the -11dB of the standard planar. The measured return loss was much better than expected. The measured gain was slightly less than the simulation stated. The radiation pattern depicts the butterfly shaped E pattern that was measured which is in good agreement with the simulated pattern.

**Table 5-4:** Measured and simulated results from fabricated planar with dish.

| Ground Plane Type Planar Dish | $S_{11}$ at Resonant Frequency (GHz) | $Z_{in}$ (Ω) | -10dB Bandwidth (%) | Gain (dB) |
|---|---|---|---|---|
| Measured | -64dB at 1.3 | 49.5+.1j | 24 | .7 |
| Simulated | -28dB at 1.3 | 40+1j | 20 | .9 |

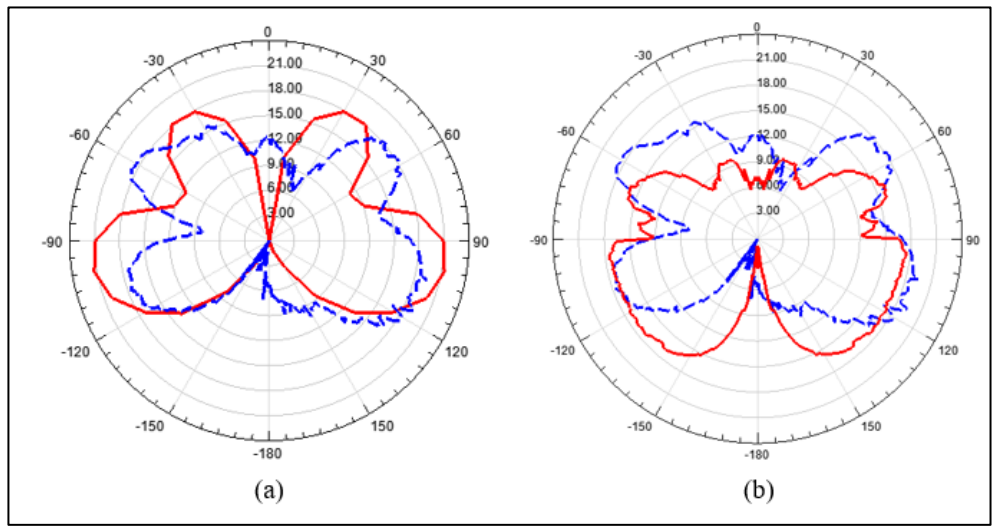

**Figure 5.6:** (a) Measured E field from fabricated planar with dish (Blue, Dash) and simulated (Red, Solid), (b) measured E field from fabricated planar with dish (Blue, Dash) and measured standard planar (Red, Solid).

## 5.2.2 Planar with Horn

The fabrication of the planar ground with the horn is very similar to the fabrication of the planar dish. The horn was 3-D printed with the optimized dimensions, 25mm in length, width, and height. The horn was electroplated with a smaller amount of current, for the same amount of time as the dish, as they are similar in size, but the horn has less surface area. A current of .35Amps was used over 2 hours to accumulate 0.9E-1mm of copper on the surface of the 3-D printed horn. The fabricated planar with horn ground can be seen in Figure 5.7. The measured results compared to the simulated results can be seen in Table 5-5, and the measured radiation pattern is shown in Figure 5.8.

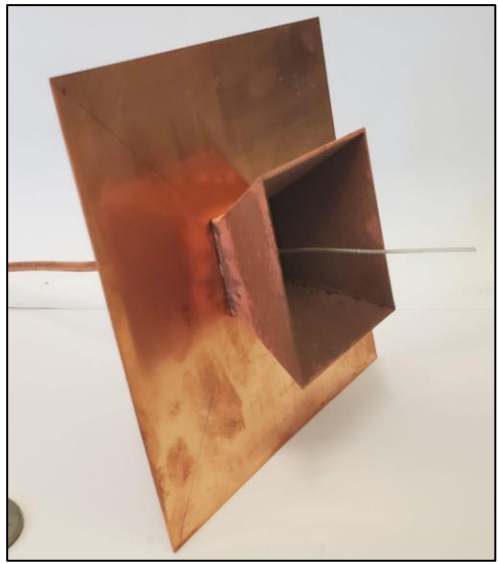

**Figure 5.7:** Fabricated planar with horn ground.

**Table 5-5:** Measured and simulated results of antenna with planar ground with horn.

| Ground Plane Type Planar Horn | $S_{11}$ at Resonant Frequency (GHz) | $Z_{in}$ (Ω) | -10dB Bandwidth (%) | Gain (dB) |
|---|---|---|---|---|
| Measured | -30dB at 1.2 | 49+2.7j | 32 | 3 |
| Simulated | -31dB at 1.345 | 32.3+4.8j | 20 | 2 |

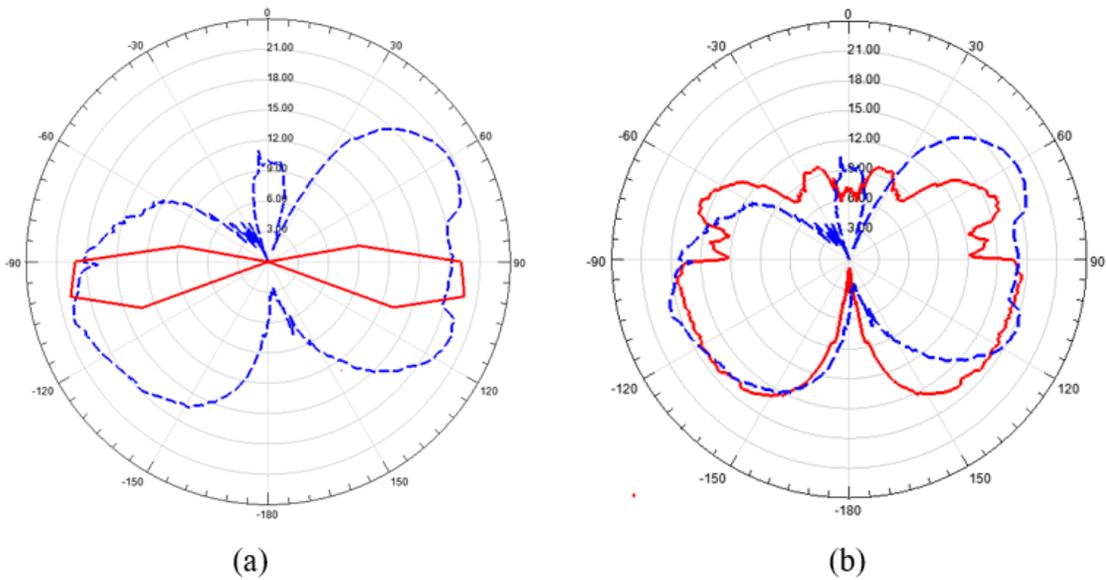

**Figure 5.8:** (a) Measured E field from fabricated planar with horn (Blue, Dash) and simulated (Red, Solid), (b) measured E field from fabricated planar with horn (Blue, Dash) and measured standard planar (Red, Solid).

The horn showed improved antenna performance. The EM simulation predicted a larger bandwidth over the standard planar, and the fabricated horn produced an even greater bandwidth improvement of up to 32%. This is a drastic improvement from the standard planar which had 6% bandwidth. The simulation predicted that the resonant frequency increased to 1.345GHz rather than 1.3GHz. However, the fabricated antenna had a resonant frequency of 1.2GHz. This may be due to fabrication and assembly tolerances. The simulated radiation pattern was exaggerated in how narrow it would be, but it did in fact create only two major lobes directly other sides rather than the butterfly shaped radiation pattern of the standard planar.

### 5.2.3 Different Size Spheres

As shown in the simulations, multiple spheres with different radii were tested. The $1/8\lambda$ and $3/8\lambda$ were fabricated. The $3/8\lambda$ was the maximum size that could fit on the 3-D print bed. Furthermore, electroplating the $3/8\lambda$ sphere, a much larger container was required than what was used for the standard sphere. A 6 gallon chemically safe bucket was used. The electroplating time and current for each sphere were found by taking the ratio of surface area relative to the standard sphere. The $1/8\lambda$ sphere was electroplated at .5Amps for 2hours while the $3/8\lambda$ sphere was plated at 2.2Amps for 4 hours. The fabricated $1/8\lambda$ and $3/8\lambda$ grounds mounted on the coaxial line are illustrated in Figure 5.9. Table 5-6 and 5-7 show the measured results from the $1/8\lambda$ and $3/8\lambda$ spheres, respectively. Figure 5.10 and 5.11 show the $1/8\lambda$ and $3/8\lambda$ radiation patterns compared to their EM simulation.

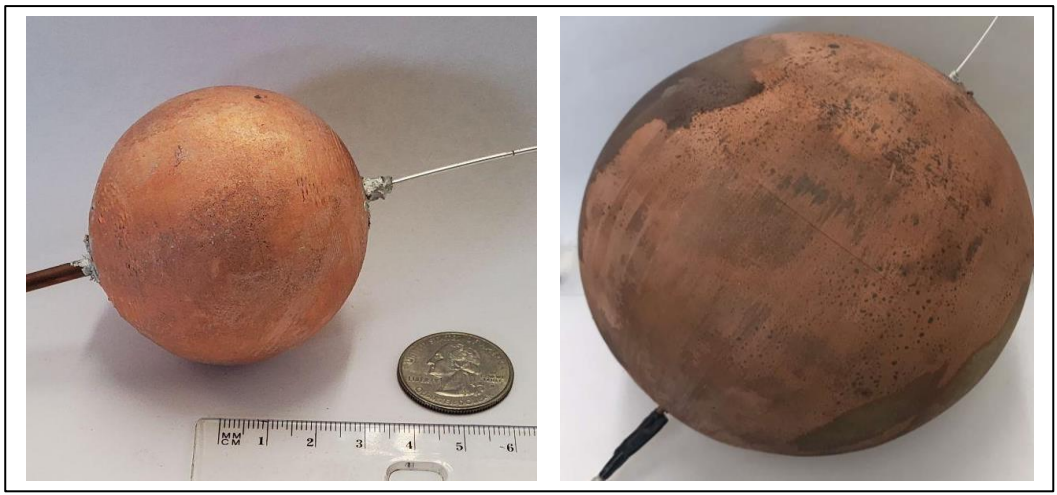

**Figure 5.9:** $1/8\lambda$ (Left) and $3/8\lambda$ (Right) fabricated ground planes.

**Table 5-6:** Measured and simulated results for the 1/8λ sphere.

| Ground Plane Type 1/8λ Sphere | $S_{11}$ at Resonant Frequency (GHz) | $Z_{in}$ (Ω) | -10dB Bandwidth (%) | Gain (dB) |
|---|---|---|---|---|
| Measured | -32dB at 1.2 | 50.4-2j | 14 | 3 |
| Simulated | -40dB at 1.22 | 61-20j | 17 | 2.8 |

**Table 5-7:** Measured and simulated results for the 3/8λ sphere.

| Ground Plane Type 3/8λ Sphere | $S_{11}$ at Resonant Frequency (GHz) | $Z_{in}$ (Ω) | -10dB Bandwidth (%) | Gain (dB) |
|---|---|---|---|---|
| Measured | -35dB at 1.27 | 53.4-1.1j | 21 | 2.5 |
| Simulated | -26dB at 1.27 | 52.9-11.6j | 17 | 2.8 |

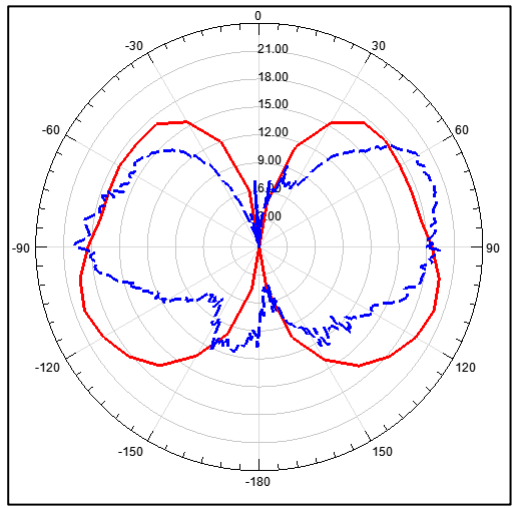

**Figure 5.10:** 1/8λ radiation pattern (Blue, Dash) compared to simulated (Red, Solid).

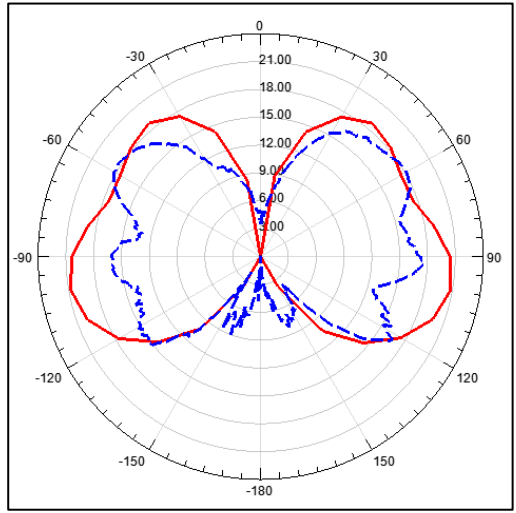

**Figure 5.11:** 3/8λ radiation pattern (Blue, Dash) compared to simulated (Red, Solid).

From Table 5-6, the 1/8λ fabricated ground plane measured response closely aligns with the simulated response. The $S_{11}$ return loss and bandwidth were slightly less than expected while the measured gain was slightly higher than the simulation. The radiation pattern shown in Figure 5.10 shows that the radiation pattern for the 1/8λ sphere was more focused towards the sides than towards the front or rear of the polar plot.

Table 5-7 shows the 3/8λ sphere outperformed the EM simulation. The $S_{11}$ return loss and bandwidth were larger than expected. The radiation pattern shown in Figure 5.11 shows a radiation pattern that matches the simulation well. The pattern is more oblong towards the front and rear, with small minor lobes slightly protruding from the sides.

### 5.2.4 Slotted Sphere

The slotted sphere is the same size as the standard sphere, but has the electroplating masked off from certain sections of the ground, exposing ABS strips along the ground plane. Painter's tape was trimmed using an exacto knife into the proper dimensions for the slotted strips, 55mm in length by 3.6mm in width. The tape was applied prior to spraying the base conductor seed layer of nickel spray on the ABS. The fabricated slotted sphere ground plane mounted on the coaxial line can be seen in Figure 5.12. The measured results compared to the simulated can be seen in Table 5-8 and the radiation pattern can be seen in Figure 5.13.

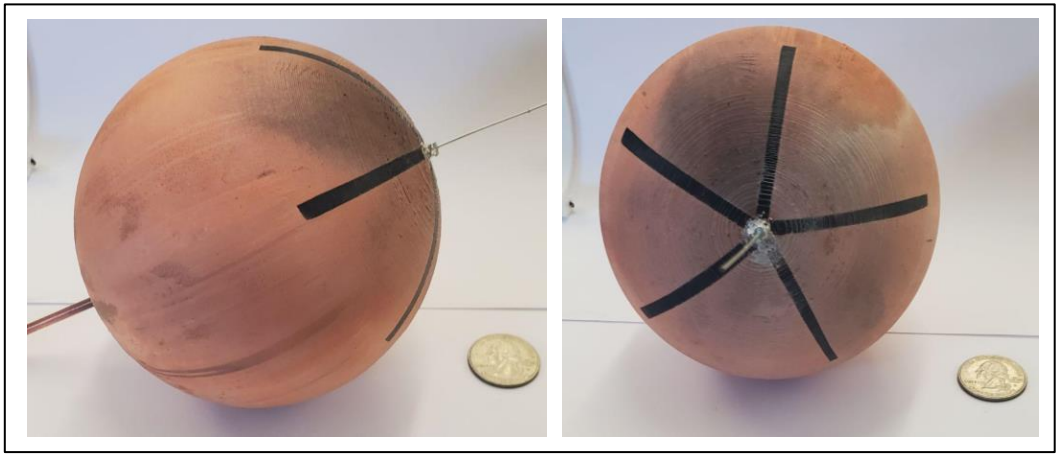

**Figure 5.12:** Fabricated slotted sphere.

**Table 5-8:** Measured and simulated results from fabricated slotted sphere.

| Ground Plane Type Slotted Sphere | $S_{11}$ at Resonant Frequency (GHz) | $Z_{in}$ ($\Omega$) | -10dB Bandwidth (%) | Gain (dB) |
|---|---|---|---|---|
| Measured | -18dB at 1.3 | 48-16.2i | 15 | 2.9 |
| Simulated | -19dB at 1.2 | 45.2-7j | 20 | 2.6 |

As shown in Table 5-8 shows that the slotted sphere performed reasonably well. The $S_{11}$ return loss matches the simulation, as does the gain. However, the measured bandwidth was only 15% while the simulation results predicted 20%. Furthermore, the measured slotted sphere impedance had a larger capacitive imaginary component than expected from the simulation. The radiation pattern shown in Figure 5.13 is in good agreement with the simulated E radiation pattern.

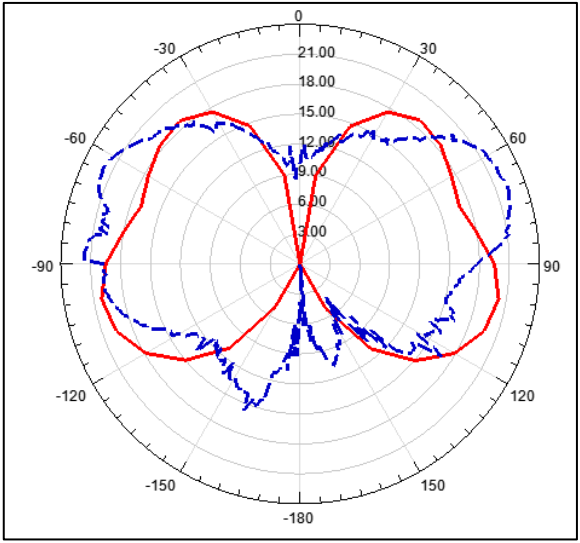

**Figure 5.13:** Slotted sphere radiation pattern (Blue, Dash) compared to simulated (Red, Solid).

## 5.2.5 Edge-Mounted Sphere

The edge-mounted sphere is the same size as the standard sphere, the monopole is just located on the "edge" of the sphere, rather than the center when viewed from the top. Figure 5.14 displays the fabricated edge-mounted sphere.

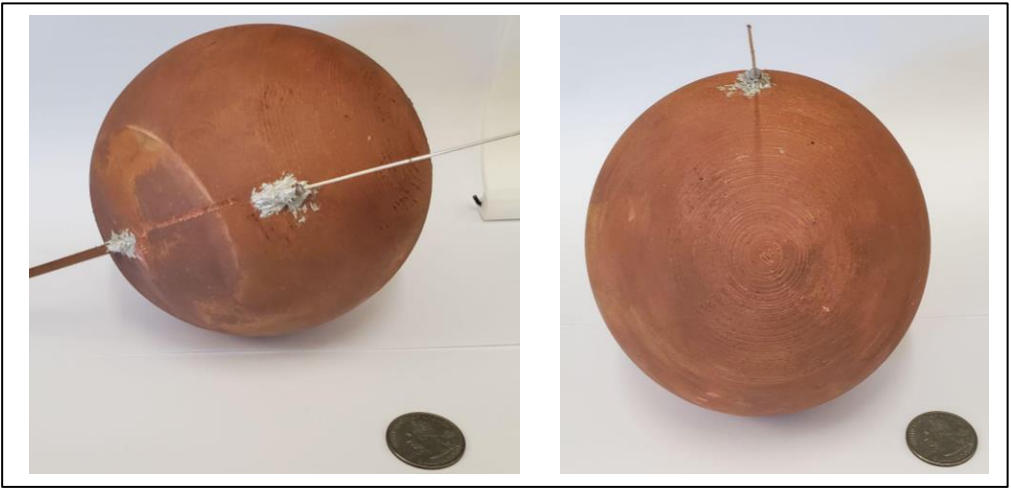

**Figure 5.14:** Fabricated edge-mounted ground with monopole antenna.

**Table 5-9:** Measured and simulated results from fabricated edge-mounted sphere.

| Edge-Mounted Sphere | $S_{11}$ at Resonant Frequency (GHz) | $Z_{in}$ (Ω) | -10dB Bandwidth (%) | Gain (dB) |
|---|---|---|---|---|
| Measured | -22dB at 1.27 | 46.7+.8j | 17 | 2 |
| Simulated | -36dB at 1.34 | 51+1.56j | 11 | 2.8 |

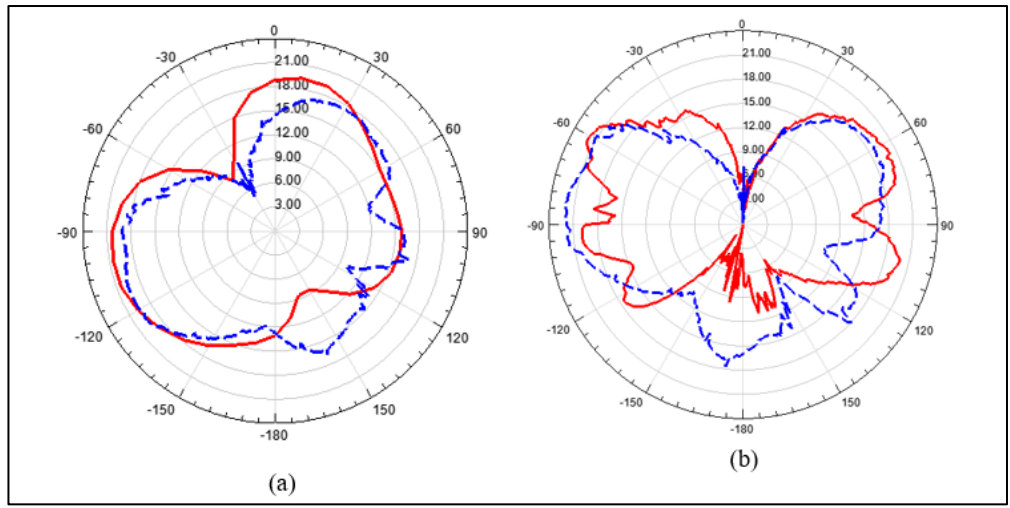

**Figure 5.15:** (a) Edge-mounted radiation pattern (Blue, Dash) compared to simulated (Red, Solid), (b) measured E field from fabricated edge-mounted (Blue, Dash) and measured standard planar (Red, Solid).

The edge-mounted antenna performed better than expected in terms of bandwidth. The measured bandwidth was 17% while the simulated bandwidth only predicted 11%. The $S_{11}$ return loss and gain were lower than the simulation suggested. A gain of 2 dB was measured when the sphere was positioned with the antenna facing the target. The radiation pattern measured seen in Figure 5.14 matches the simulated radiation pattern almost exactly. The edge location of the monopole led to a directive radiation pattern and gain on the spherical ground plane.

## 5.2.6 Sphere Composed of Fins

The final fabricated sphere was the sphere composed of fins. This geometry is rather abstract and seems to produce improvements for the monopole. The Fin sphere proved to be difficult to electroplate properly. Upon the first electroplating test, all the copper was attracted to the outer sections of the fins, leaving the inner crevasses not plated. This is because the current flow carrying the copper ions wanted to take the shortest path possible to ground, the outer sections of the fins are closest to the positively charged copper plates. To try to combat this, the grounded wire that is mounted to the surface of the part was placed deep into the crevasses of the fins, to force copper ions into the regions between the fins. The current was increased to 1.85 Amps due to the increased surface area of the part compared to the standard sphere. Figure 5.16 shows the partially coated electroplated fins, even with the ground wire in the crevasse.

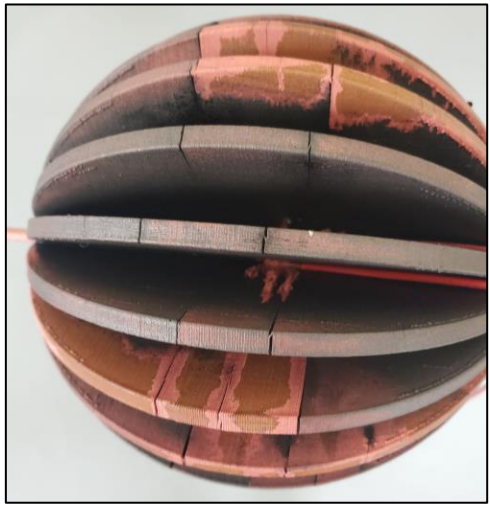

**Figure 5.16:** Partially coated fin sphere.

In another attempt to electroplate this complex geometry, the surface was sprayed with a more conductive base seed layer of the MG Chemicals Silver Coated Copper Conductive Paint.

After electroplating with this more conductive coating as the base layer, the electroplating succeeded in evenly coating the entire surface. This evenly plated fin sphere surface can be seen in Figure 5.17. Table 5-10 shows the measured results with the simulated. Figure 5.18 shows the measured radiation pattern compared to the simulated.

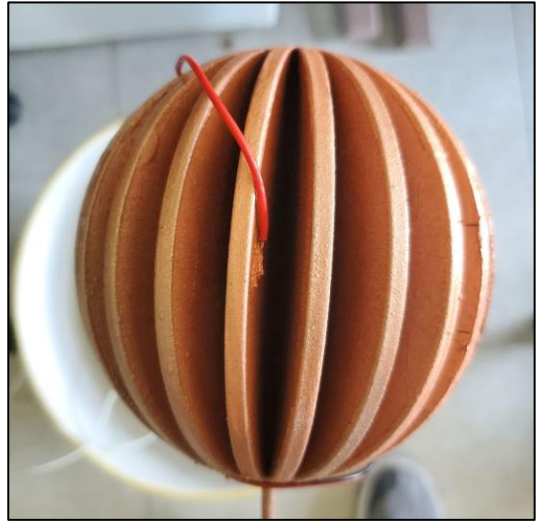

**Figure 5.17:** Successfully electroplated fin sphere ground.

**Table 5-10:** Measured and simulated results from fabricated fin sphere.

| Ground Plane Type Fin Sphere | $S_{11}$ at Resonant Frequency (GHz) | $Z_{in}$ ($\Omega$) | -10dB Bandwidth (%) | Gain (dB) |
|---|---|---|---|---|
| Measured | -34dB at 1.3 | 52-.1j | 18 | 4 |
| Simulated | -16dB at 1.3 | 62-12j | 24 | 3 |

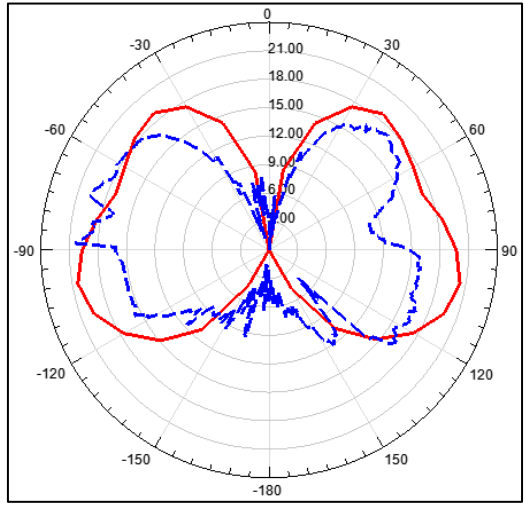

**Figure 5.18**: Fin sphere radiation pattern (Blue, Dash) compared to simulated (Red, Solid).

The fin sphere shows improvement over the standard sphere as well as some improvement over what the simulated predicted. The $S_{11}$ return loss was over double the value the simulation stated at -34dB. However, the bandwidth was narrower than what was expected. The measured fin sphere had a better match and larger gain than the simulation presented. Furthermore, the radiation pattern shown in Figure 5.17 is in good agreement with the simulation.

## 5.3 Discussion

### 5.3.1 Fabricated Planar Improvements

All the fabricated altered planar grounds drastically improved the performance of the standard planar ground. Both the dish and horn increased not only the bandwidth but also the return loss and match compared to the actual fabricated standard planar. Table 5-11 shows the actual fabricated altered planar grounds in comparison to the standard fabricated ground.

**Table 5-11:** Measured results of fabricated altered planar grounds and the standard planar.

| Ground Plane Type | $S_{11}$ at Resonant Frequency (GHz) | $Z_{in}$ ($\Omega$) | -10dB Bandwidth (%) | Gain (dB) |
|---|---|---|---|---|
| Measured Standard Planar | -10dB at 1.2 | 27.5-2.7j | 6 | 2 |
| Measured Planar Dish | -64dB at 1.28 | 49.5+.1j | 24 | .7 |
| Measured Planar Horn | -30dB at 1.2 | 49+2.7j | 32 | 3 |

With only the addition of the 3-D printed, electroplated copper dish of optimized size, the bandwidth of the planar ground plane went from 6% to 24%, and the $S_{11}$ return loss went from only -10dB to -64dB. However, for the planar dish, the gain decreased. With adding the 3-D printed electroplated horn, the bandwidth was increased even more to 32% and the gain was increased to 3dB. Both small additions to the standard planar geometry proved to increase performance drastically for the monopole antenna, which was verified in simulation and in fabrication.

### 5.3.2 Different Sized Fabricated Spheres

Just simply altering the radius of the standard sphere seemed to produce trends as the radius gets larger. The different radius spheres that were fabricated are the smaller 1/8λ and the larger 3/8λ radius spheres. 3/8λ was the maximum that the 3-D printer electroplating station could hold with the object in 1 piece. A table showing all the fabricated spheres of different radii can be seen in Table 5-12.

**Table 5-12:** Measured results of fabricated 1/8λ, 1/4λ and 3/8λ spheres.

| Ground Plane Type (radius) | $S_{11}$ at Resonant Frequency (GHz) | $Z_{in}$ (Ω) | -10dB Bandwidth (%) | Gain (dB) |
|---|---|---|---|---|
| Measured 1/8λ | -32dB at 1.2 | 50.4-2j | 14 | 3 |
| Measured 1/4λ (Standard Sphere) | -37dB at 1.28 | 50-2.4j | 16 | 3.3 |
| Measured 1/4λ | -35dB at 1.27 | 53.4-1.1j | 21 | 2.5 |

It seems that as the radius increased, the bandwidth also increased as well. The simulation predicted that the bandwidth would decrease as the radius increased, which is contradictory to the fabricated results shown in Table 5-12. The gain however does follow the simulation's prediction of it decreasing as the radius increases.

### 5.2.3 Fabricated Edge-Mounted Sphere

The fabricated edge-mounted sphere proved to produce a directional radiation pattern and gain by simply altering the location of the antenna's mount location on the ground plane. This was shown in the simulation and in the actual fabrication of the edge-mounted sphere. The measured radiation patterns of the edge-mounted and standard sphere can be seen overlaid with each other in Figure 5.19. Table 5-13 displays the measured characteristics of the fabricated edge-mount and standard sphere.

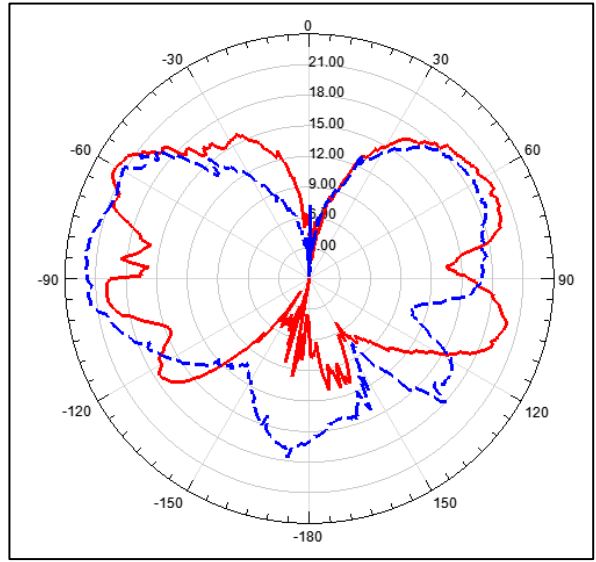

**Figure 5.19:** Measured edge-mounted E radiation pattern (Blue, Dash) compared to measured standard sphere E (Red, Solid).

**Table 5-13:** Measured results of fabricated edge-mounted and standard sphere.

| Ground Plane Type | $S_{11}$ at Resonant Frequency (GHz) | $Z_{in}$ ($\Omega$) | -10dB Bandwidth (%) | Gain (dB) |
|---|---|---|---|---|
| Measured Edge-Mounted | -22dB at 1.27 | 46.7+.8j | 17 | 2 (Directional) |
| Measured Standard Sphere | -37dB at 1.28 | 50-2.4j | 16 | 3.3 |

As shown, the fabricated edge-mounted ground plane didn't seem to improve on the $S_{11}$ return loss, bandwidth, or match. The feature that makes the edge-mounted ground unique is that the radiation pattern can be made directional without altering the actual ground plane, just the antenna mount location. Shown in Figure 5.18, the left major lobe is larger in size while the right size lobe stays the same size as the standard sphere. This can be especially useful when communicating in a RF application with a known destination target to focus as much signal to the destination as possible.

### 5.2.4 Fabricated Sphere composed of Fins

The fabricated sphere composed of fins did show some improvement over the standard sphere ground in simulation. Table 5-14 shows the measured results of the fabricated fin sphere and standard sphere.

**Table 5-14:** Measured results of fabricated fin sphere and standard sphere.

| Ground Plane Type | $S_{11}$ at Resonant Frequency (GHz) | $Z_{in}$ ($\Omega$) | -10dB Bandwidth (%) | Gain (dB) |
|---|---|---|---|---|
| Measured Fin Sphere | -34dB at 1.28 | 52-.1j | 18 | 4 |
| Measured Standard Sphere | -37dB at 1.28 | 50-2.4j | 16 | 3.3 |

As shown in the figure above, the fabricated fin sphere did outperform the standard sphere in almost every measurement. The measured bandwidth and gain were both higher than the fabricated standard sphere. The imaginary component of the match improved. The $S_{11}$ return loss was a slightly lower numerical value than the standard sphere, but not a significant decrease by any means. The fabricated fin sphere did perform better than the standard fin sphere which shows that complex geometry can be successfully 3-D printed with post processing steps to produce functional RF components, in this case, a complex ground plane.

# Chapter 6

## 6.1 Conclusion

Modifications can be made to standard planar and spherical ground planes to modify antenna performance using 3-D printing. Multiple designs for each of the planar and spherical grounds were simulated, fabricated, and measured. Complex geometry was also introduced to display capabilities of 3-D printed RF components. The planar ground showed great improvement with the additions of a reflective dish and horn that was attached to the ground plane. The $S_{11}$ return loss, gain and operating bandwidth dramatically increased in performance with these small 3-D printed additions to the ground plane. Moreover, the spherical ground plane exhibited interesting trends found as the radius of the sphere changes. The actual fabricated multi-radius spheres show an increase in bandwidth and a decrease in gain as the radius increases. Moreover, the mounting location on a spherical ground plane can modify the radiation characteristics of the monopole antenna. A complex spherical geometry was also displayed to feature the capabilities of 3-D printing in RF applications.

### 6.1.1 Future Work

There were many simulated models that displayed potential in modifying the antennas performance but couldn't be fabricated due to time constraints. This includes the planar cone which shifted the resonant frequency drastically by more than 1GHz. It would be interesting to see if this occurs with a real fabricated model.

Furthermore, mechanical reliability of the 3-D printed ground planes could be researched to examine how RF characteristics would change if the ground plane designs were in extreme temperatures, under stress, or even extreme humidity. Moreover, it would be interesting to take an alternative approach and take each modified ground plane and vary the antenna length to optimize and tune the antenna for the modified ground plane.